
%
\catcode`@=11 
%
%
%

\font\fourteenrm=cmr10 scaled\magstep2
\font\twelverm=cmr10 scaled\magstep1
\font\ninerm=cmr9            \font\sixrm=cmr6

\font\fourteenbf=cmbx10 scaled\magstep2
\font\twelvebf=cmbx10 scaled\magstep1
\font\ninebf=cmbx9            \font\sixbf=cmbx6
\font\seventeeni=cmmi10 scaled\magstep3     \skewchar\seventeeni='177
\font\fourteeni=cmmi10 scaled\magstep2      \skewchar\fourteeni='177
\font\twelvei=cmmi10 scaled\magstep1        \skewchar\twelvei='177
\font\ninei=cmmi9                           \skewchar\ninei='177
\font\sixi=cmmi6                            \skewchar\sixi='177
\font\seventeensy=cmsy10 scaled\magstep3    \skewchar\seventeensy='60
\font\fourteensy=cmsy10 scaled\magstep2     \skewchar\fourteensy='60
\font\twelvesy=cmsy10 scaled\magstep1       \skewchar\twelvesy='60
\font\ninesy=cmsy9                          \skewchar\ninesy='60
\font\sixsy=cmsy6                           \skewchar\sixsy='60

\font\fourteenex=cmex10 scaled\magstep2
\font\twelveex=cmex10 scaled\magstep1

\font\fourteensl=cmsl10 scaled\magstep2
\font\twelvesl=cmsl10 scaled\magstep1
\font\ninesl=cmsl9

\font\fourteenit=cmti10 scaled\magstep2
\font\twelveit=cmti10 scaled\magstep1
\font\twelvett=cmtt10 scaled\magstep1
\font\twelvecp=cmcsc10 scaled\magstep1
\font\tencp=cmcsc10
\newfam\cpfam
%
%
\newcount\f@ntkey            \f@ntkey=0
\def\samef@nt{\relax \ifcase\f@ntkey \rm \or\oldstyle \or\or
         \or\it \or\sl \or\bf \or\tt \or\caps \fi }
\def\fourteenpoint{\relax
    \textfont0=\fourteenrm          \scriptfont0=\tenrm
    \scriptscriptfont0=\sevenrm
     \def\rm{\fam0 \fourteenrm \f@ntkey=0 }\relax
    \textfont1=\fourteeni           \scriptfont1=\teni
    \scriptscriptfont1=\seveni
     \def\oldstyle{\fam1 \fourteeni\f@ntkey=1 }\relax
    \textfont2=\fourteensy          \scriptfont2=\tensy
    \scriptscriptfont2=\sevensy
    \textfont3=\fourteenex     \scriptfont3=\fourteenex
    \scriptscriptfont3=\fourteenex
    \def\it{\fam\itfam \fourteenit\f@ntkey=4 }\textfont\itfam=\fourteenit
    \def\sl{\fam\slfam \fourteensl\f@ntkey=5 }\textfont\slfam=\fourteensl
    \scriptfont\slfam=\tensl
    \def\bf{\fam\bffam \fourteenbf\f@ntkey=6 }\textfont\bffam=\fourteenbf
    \scriptfont\bffam=\tenbf     \scriptscriptfont\bffam=\sevenbf
    \def\tt{\fam\ttfam \twelvett \f@ntkey=7 }\textfont\ttfam=\twelvett
    \h@big=11.9\p@{} \h@Big=16.1\p@{} \h@bigg=20.3\p@{} \h@Bigg=24.5\p@{}
    \def\caps{\fam\cpfam \twelvecp \f@ntkey=8 }\textfont\cpfam=\twelvecp
    \setbox\strutbox=\hbox{\vrule height 12pt depth 5pt width\z@}
    \samef@nt}
\def\twelvepoint{\relax
    \textfont0=\twelverm          \scriptfont0=\ninerm
    \scriptscriptfont0=\sixrm
     \def\rm{\fam0 \twelverm \f@ntkey=0 }\relax
    \textfont1=\twelvei           \scriptfont1=\ninei
    \scriptscriptfont1=\sixi
     \def\oldstyle{\fam1 \twelvei\f@ntkey=1 }\relax
    \textfont2=\twelvesy          \scriptfont2=\ninesy
    \scriptscriptfont2=\sixsy
    \textfont3=\twelveex          \scriptfont3=\twelveex
    \scriptscriptfont3=\twelveex
    \def\it{\fam\itfam \twelveit \f@ntkey=4 }\textfont\itfam=\twelveit
    \def\sl{\fam\slfam \twelvesl \f@ntkey=5 }\textfont\slfam=\twelvesl
    \scriptfont\slfam=\ninesl
    \def\bf{\fam\bffam \twelvebf \f@ntkey=6 }\textfont\bffam=\twelvebf
    \scriptfont\bffam=\ninebf     \scriptscriptfont\bffam=\sixbf
    \def\tt{\fam\ttfam \twelvett \f@ntkey=7 }\textfont\ttfam=\twelvett
    \h@big=10.2\p@{}
    \h@Big=13.8\p@{}
    \h@bigg=17.4\p@{}
    \h@Bigg=21.0\p@{}
    \def\caps{\fam\cpfam \twelvecp \f@ntkey=8 }\textfont\cpfam=\twelvecp
    \setbox\strutbox=\hbox{\vrule height 10pt depth 4pt width\z@}
    \samef@nt}
\def\tenpoint{\relax
    \textfont0=\tenrm          \scriptfont0=\sevenrm
    \scriptscriptfont0=\fiverm
    \def\rm{\fam0 \tenrm \f@ntkey=0 }\relax
    \textfont1=\teni           \scriptfont1=\seveni
    \scriptscriptfont1=\fivei
    \def\oldstyle{\fam1 \teni \f@ntkey=1 }\relax
    \textfont2=\tensy          \scriptfont2=\sevensy
    \scriptscriptfont2=\fivesy
    \textfont3=\tenex          \scriptfont3=\tenex
    \scriptscriptfont3=\tenex
    \def\it{\fam\itfam \tenit \f@ntkey=4 }\textfont\itfam=\tenit
    \def\sl{\fam\slfam \tensl \f@ntkey=5 }\textfont\slfam=\tensl
    \def\bf{\fam\bffam \tenbf \f@ntkey=6 }\textfont\bffam=\tenbf
    \scriptfont\bffam=\sevenbf     \scriptscriptfont\bffam=\fivebf
    \def\tt{\fam\ttfam \tentt \f@ntkey=7 }\textfont\ttfam=\tentt
    \def\caps{\fam\cpfam \tencp \f@ntkey=8 }\textfont\cpfam=\tencp
    \setbox\strutbox=\hbox{\vrule height 8.5pt depth 3.5pt width\z@}
    \samef@nt}
%
%
%
%
\newdimen\h@big  \h@big=8.5\p@
\newdimen\h@Big  \h@Big=11.5\p@
\newdimen\h@bigg  \h@bigg=14.5\p@
\newdimen\h@Bigg  \h@Bigg=17.5\p@
\def\big#1{{\hbox{$\left#1\vbox to\h@big{}\right.\n@space$}}}
\def\Big#1{{\hbox{$\left#1\vbox to\h@Big{}\right.\n@space$}}}
\def\bigg#1{{\hbox{$\left#1\vbox to\h@bigg{}\right.\n@space$}}}
\def\Bigg#1{{\hbox{$\left#1\vbox to\h@Bigg{}\right.\n@space$}}}
%
%
%
\normalbaselineskip = 20pt plus 0.2pt minus 0.1pt
\normallineskip = 1.5pt plus 0.1pt minus 0.1pt
\normallineskiplimit = 1.5pt
\newskip\normaldisplayskip
\normaldisplayskip = 20pt plus 5pt minus 10pt
\newskip\normaldispshortskip
\normaldispshortskip = 6pt plus 5pt
\newskip\normalparskip
\normalparskip = 6pt plus 2pt minus 1pt
\newskip\skipregister
\skipregister = 5pt plus 2pt minus 1.5pt
\newif\ifsingl@    \newif\ifdoubl@
\newif\iftwelv@    \twelv@true
\def\singlespace{\singl@true\doubl@false\spaces@t}
\def\doublespace{\singl@false\doubl@true\spaces@t}
\def\normalspace{\singl@false\doubl@false\spaces@t}
\def\Tenpoint{\tenpoint\twelv@false\spaces@t}
\def\Twelvepoint{\twelvepoint\twelv@true\spaces@t}
\def\spaces@t{\relax%
 \iftwelv@ \ifsingl@\subspaces@t3:4;\else\subspaces@t29:31;\fi%
 \else \ifsingl@\subspaces@t3:5;\else\subspaces@t4:5;\fi \fi%
 \ifdoubl@ \multiply\baselineskip by 5%
 \divide\baselineskip by 4 \fi \unskip}
\def\subspaces@t#1:#2;{
      \baselineskip = \normalbaselineskip
      \multiply\baselineskip by #1 \divide\baselineskip by #2
      \lineskip = \normallineskip
      \multiply\lineskip by #1 \divide\lineskip by #2
      \lineskiplimit = \normallineskiplimit
      \multiply\lineskiplimit by #1 \divide\lineskiplimit by #2
      \parskip = \normalparskip
      \multiply\parskip by #1 \divide\parskip by #2
      \abovedisplayskip = \normaldisplayskip
      \multiply\abovedisplayskip by #1 \divide\abovedisplayskip by #2
      \belowdisplayskip = \abovedisplayskip
      \abovedisplayshortskip = \normaldispshortskip
      \multiply\abovedisplayshortskip by #1
        \divide\abovedisplayshortskip by #2
      \belowdisplayshortskip = \abovedisplayshortskip
      \advance\belowdisplayshortskip by \belowdisplayskip
      \divide\belowdisplayshortskip by 2
      \smallskipamount = \skipregister
      \multiply\smallskipamount by #1 \divide\smallskipamount by #2
      \medskipamount = \smallskipamount \multiply\medskipamount by 2
      \bigskipamount = \smallskipamount \multiply\bigskipamount by 4 }
\def\normalbaselines{ \baselineskip=\normalbaselineskip
   \lineskip=\normallineskip \lineskiplimit=\normallineskip
   \iftwelv@\else \multiply\baselineskip by 4 \divide\baselineskip by 5
     \multiply\lineskiplimit by 4 \divide\lineskiplimit by 5
     \multiply\lineskip by 4 \divide\lineskip by 5 \fi }
\Twelvepoint  
\interlinepenalty=50
\interfootnotelinepenalty=5000
\predisplaypenalty=9000
\postdisplaypenalty=500
\hfuzz=1pt
\vfuzz=0.2pt
%
%
%
\def\pagecontents{
   \ifvoid\topins\else\unvbox\topins\vskip\skip\topins\fi
   \dimen@ = \dp255 \unvbox255
   \ifvoid\footins\else\vskip\skip\footins\footrule\unvbox\footins\fi
   \ifr@ggedbottom \kern-\dimen@ \vfil \fi }
\def\makeheadline{\vbox to 0pt{ \skip@=\topskip
      \advance\skip@ by -12pt \advance\skip@ by -2\normalbaselineskip
      \vskip\skip@ \line{\vbox to 12pt{}\the\headline} \vss
      }\nointerlineskip}
\def\makefootline{\baselineskip = 1.5\normalbaselineskip
                 \line{\the\footline}}
\newif\iffrontpage
\newif\ifletterstyle
\newif\ifp@genum
\def\nopagenumbers{\p@genumfalse}
\def\pagenumbers{\p@genumtrue}
\pagenumbers
\newtoks\paperheadline
\newtoks\letterheadline
\newtoks\letterfrontheadline
\newtoks\lettermainheadline
\newtoks\paperfootline
\newtoks\letterfootline
\newtoks\date
\footline={\ifletterstyle\the\letterfootline\else\the\paperfootline\fi}
\paperfootline={\hss\iffrontpage\else\ifp@genum\tenrm\folio\hss\fi\fi}
\letterfootline={\hfil}
\headline={\ifletterstyle\the\letterheadline\else\the\paperheadline\fi}
\paperheadline={\hfil}
\letterheadline{\iffrontpage\the\letterfrontheadline
     \else\the\lettermainheadline\fi}
\lettermainheadline={\rm\ifp@genum page \ \folio\fi\hfil\the\date}
\def\monthname{\relax\ifcase\month 0/\or January\or February\or
   March\or April\or May\or June\or July\or August\or September\or
   October\or November\or December\else\number\month/\fi}
\date={\monthname\ \number\day, \number\year}
\countdef\pagenumber=1  \pagenumber=1
\def\advancepageno{\global\advance\pageno by 1
   \ifnum\pagenumber<0 \global\advance\pagenumber by -1
    \else\global\advance\pagenumber by 1 \fi \global\frontpagefalse }
\def\folio{\ifnum\pagenumber<0 \romannumeral-\pagenumber
           \else \number\pagenumber \fi }
\def\footrule{\dimen@=\prevdepth\nointerlineskip
   \vbox to 0pt{\vskip -0.25\baselineskip \hrule width 0.35\hsize \vss}
   \prevdepth=\dimen@ }
\newtoks\foottokens
\foottokens={\Tenpoint\singlespace}
\newdimen\footindent
\footindent=24pt
\def\vfootnote#1{\insert\footins\bgroup  \the\foottokens
   \interlinepenalty=\interfootnotelinepenalty \floatingpenalty=20000
   \splittopskip=\ht\strutbox \boxmaxdepth=\dp\strutbox
   \leftskip=\footindent \rightskip=\z@skip
   \parindent=0.5\footindent \parfillskip=0pt plus 1fil
   \spaceskip=\z@skip \xspaceskip=\z@skip
   \Textindent{$ #1 $}\footstrut\futurelet\next\fo@t}
\def\Textindent#1{\noindent\llap{#1\enspace}\ignorespaces}
\def\footnote#1{\attach{#1}\vfootnote{#1}}

\let\footsymbol=\star
\newcount\lastf@@t           \lastf@@t=-1
\newcount\footsymbolcount    \footsymbolcount=0
\newif\ifPhysRev
\def\footsymbolgen{\relax \ifPhysRev \iffrontpage \NPsymbolgen\else
      \PRsymbolgen\fi \else \NPsymbolgen\fi
   \global\lastf@@t=\pageno \footsymbol }
\def\NPsymbolgen{\ifnum\footsymbolcount<0 \global\footsymbolcount=0\fi
   {\iffrontpage \else \advance\lastf@@t by 1 \fi
    \ifnum\lastf@@t<\pageno \global\footsymbolcount=0
     \else \global\advance\footsymbolcount by 1 \fi }
   \ifcase\footsymbolcount \fd@f\star\or \fd@f\dagger\or \fd@f\ast\or
    \fd@f\ddagger\or \fd@f\natural\or \fd@f\diamond\or \fd@f\bullet\or
    \fd@f\nabla\else \fd@f\dagger\global\footsymbolcount=0 \fi }
\def\fd@f#1{\xdef\footsymbol{#1}}
\def\PRsymbolgen{\ifnum\footsymbolcount>0 \global\footsymbolcount=0\fi
      \global\advance\footsymbolcount by -1
      \xdef\footsymbol{\sharp\number-\footsymbolcount} }
\def\space@ver#1{\let\@sf=\empty \ifmmode #1\else \ifhmode
   \edef\@sf{\spacefactor=\the\spacefactor}\unskip${}#1$\relax\fi\fi}
\def\attach#1{\space@ver{\strut^{\mkern 2mu #1} }\@sf\ }
\def\atttach#1{\space@ver{\strut{\mkern 2mu #1} }\@sf\ }
%
%
%
\newcount\chapternumber      \chapternumber=0
\newcount\sectionnumber      \sectionnumber=0
\newcount\equanumber         \equanumber=0
\let\chapterlabel=0
\newtoks\chapterstyle        \chapterstyle={\Number}
\newskip\chapterskip         \chapterskip=\bigskipamount
\newskip\sectionskip         \sectionskip=\medskipamount
\newskip\headskip            \headskip=8pt plus 3pt minus 3pt
\newdimen\chapterminspace    \chapterminspace=15pc
\newdimen\sectionminspace    \sectionminspace=10pc
\newdimen\referenceminspace  \referenceminspace=25pc
\def\chapterreset{\global\advance\chapternumber by 1
   \ifnum\the\equanumber<0 \else\global\equanumber=0\fi
   \sectionnumber=0 \makel@bel}
\def\makel@bel{\xdef\chapterlabel{%
\the\chapterstyle{\the\chapternumber}.}}
\def\sectionlabel{\number\sectionnumber \quad }
\def\alphabetic#1{\count255='140 \advance\count255 by #1\char\count255}
\def\Alphabetic#1{\count255='100 \advance\count255 by #1\char\count255}
\def\Roman#1{\uppercase\expandafter{\romannumeral #1}}
\def\roman#1{\romannumeral #1}
\def\Number#1{\number #1}
\def\unnumberedchapters{\let\makel@bel=\relax \let\chapterlabel=\relax
\let\sectionlabel=\relax \equanumber=-1 }
\def\titlestyle#1{\par\begingroup \interlinepenalty=9999
     \leftskip=0.02\hsize plus 0.23\hsize minus 0.02\hsize
     \rightskip=\leftskip \parfillskip=0pt
     \hyphenpenalty=9000 \exhyphenpenalty=9000
     \tolerance=9999 \pretolerance=9000
     \spaceskip=0.333em \xspaceskip=0.5em
     \iftwelv@\fourteenpoint\else\twelvepoint\fi
   \noindent #1\par\endgroup }
\def\spacecheck#1{\dimen@=\pagegoal\advance\dimen@ by -\pagetotal
   \ifdim\dimen@<#1 \ifdim\dimen@>0pt \vfil\break \fi\fi}
\def\chapter#1{\par \penalty-300 \vskip\chapterskip
   \spacecheck\chapterminspace
   \chapterreset \titlestyle{\chapterlabel \ #1}
   \nobreak\vskip\headskip \penalty 30000
   \wlog{\string\chapter\ \chapterlabel} }

\def\section#1{\par \ifnum\the\lastpenalty=30000\else
   \penalty-200\vskip\sectionskip \spacecheck\sectionminspace\fi
   \wlog{\string\section\ \chapterlabel \the\sectionnumber}
   \global\advance\sectionnumber by 1  \noindent
   {\caps\enspace\chapterlabel \sectionlabel #1}\par
   \nobreak\vskip\headskip \penalty 30000 }
\def\subsection#1{\par
   \ifnum\the\lastpenalty=30000\else \penalty-100\smallskip \fi
   \noindent\undertext{#1}\enspace \vadjust{\penalty5000}}

\def\undertext#1{\vtop{\hbox{#1}\kern 1pt \hrule}}
\def\APPENDIX#1#2{\par\penalty-300\vskip\chapterskip
   \spacecheck\chapterminspace \chapterreset \xdef\chapterlabel{#1}
   \titlestyle{APPENDIX #2} \nobreak\vskip\headskip \penalty 30000
   \wlog{\string\Appendix\ \chapterlabel} }
\def\Appendix#1{\APPENDIX{#1}{#1}}
\def\appendix{\APPENDIX{A}{}}
%
%
%
\def\eqname#1{\relax \ifnum\the\equanumber<0
     \xdef#1{{\rm(\number-\equanumber)}}\global\advance\equanumber by -1
    \else \global\advance\equanumber by 1
      \xdef#1{{\rm(\chapterlabel \number\equanumber)}} \fi}

\def\eqn#1{\eqno\eqname{#1}#1}

\def\eqinsert#1{\noalign{\dimen@=\prevdepth \nointerlineskip
   \setbox0=\hbox to\displaywidth{\hfil #1}
   \vbox to 0pt{\vss\hbox{$\!\box0\!$}\kern-0.5\baselineskip}
   \prevdepth=\dimen@}}
\def\sequentialequations{\globaleqnumbers}
%
%
\def\GENITEM#1;#2{\par \hangafter=0 \hangindent=#1
    \Textindent{$ #2 $}\ignorespaces}
\outer\def\newitem#1=#2;{\gdef#1{\GENITEM #2;}}
\newdimen\itemsize                \itemsize=30pt
\newitem\item=1\itemsize;
\newitem\sitem=1.75\itemsize;     
\newitem\ssitem=2.5\itemsize;     
\outer\def\newlist#1=#2&#3&#4;{\toks0={#2}\toks1={#3}%
   \count255=\escapechar \escapechar=-1
   \alloc@0\list\countdef\insc@unt\listcount     \listcount=0
   \edef#1{\par
      \countdef\listcount=\the\allocationnumber
      \advance\listcount by 1
      \hangafter=0 \hangindent=#4
      \Textindent{\the\toks0{\listcount}\the\toks1}}
   \expandafter\expandafter\expandafter
    \edef\c@t#1{begin}{\par
      \countdef\listcount=\the\allocationnumber \listcount=1
      \hangafter=0 \hangindent=#4
      \Textindent{\the\toks0{\listcount}\the\toks1}}
   \expandafter\expandafter\expandafter
    \edef\c@t#1{con}{\par \hangafter=0 \hangindent=#4 \noindent}
   \escapechar=\count255}
\def\c@t#1#2{\csname\string#1#2\endcsname}
\newlist\point=\Number&.&1.0\itemsize;
\newlist\subpoint=(\alphabetic&)&1.75\itemsize;
\newlist\subsubpoint=(\roman&)&2.5\itemsize;
\newlist\cpoint=\Roman&.&1.0\itemsize;
%

%
%
%
\newcount\referencecount     \referencecount=0
\newif\ifreferenceopen       \newwrite\referencewrite
\newtoks\rw@toks
\def\NPrefmark#1{\atttach{\rm [ #1 ] }}
\let\PRrefmark=\attach
\def\CErefmark#1{\attach{\scriptstyle  #1 ) }}
\def\refmark#1{\relax\ifPhysRev\PRrefmark{#1}\else\NPrefmark{#1}\fi}
\def\crefmark#1{\relax\CErefmark{#1}}
\def\refend{\refmark{\number\referencecount}}
\newcount\lastrefsbegincount \lastrefsbegincount=0
\def\refsend{\refmark{\count255=\referencecount
   \advance\count255 by-\lastrefsbegincount
   \ifcase\count255 \number\referencecount
   \or \number\lastrefsbegincount,\number\referencecount
   \else \number\lastrefsbegincount-\number\referencecount \fi}}
\def\crefsend{\crefmark{\count255=\referencecount
   \advance\count255 by-\lastrefsbegincount
   \ifcase\count255 \number\referencecount
   \or \number\lastrefsbegincount,\number\referencecount
   \else \number\lastrefsbegincount-\number\referencecount \fi}}
\def\refch@ck{\chardef\rw@write=\referencewrite
   \ifreferenceopen \else \referenceopentrue
   \immediate\openout\referencewrite=referenc.texauxil \fi}
%
{\catcode`\^^M=\active 
  \gdef\obeyendofline{\catcode`\^^M\active \let^^M\ }}%
%
{\catcode`\^^M=\active 
  \gdef\ignoreendofline{\catcode`\^^M=5}}
{\obeyendofline\gdef\rw@start#1{\def\t@st{#1} \ifx\t@st\blankend%
\endgroup \@sf \relax \else \ifx\t@st\bl@nkend \endgroup \@sf \relax%
\else \rw@begin#1
\backtotext
\fi \fi } }
{\obeyendofline\gdef\rw@begin#1
{\def\n@xt{#1}\rw@toks={#1}\relax%
\rw@next}}
\def\blankend{}
{\obeylines\gdef\bl@nkend{
}}
\newif\iffirstrefline  \firstreflinetrue
\def\rwr@teswitch{\ifx\n@xt\blankend \let\n@xt=\rw@begin %
 \else\iffirstrefline \global\firstreflinefalse%
\immediate\write\rw@write{\noexpand\obeyendofline \the\rw@toks}%
\let\n@xt=\rw@begin%
      \else\ifx\n@xt\rw@@d \def\n@xt{\immediate\write\rw@write{%
        \noexpand\ignoreendofline}\endgroup \@sf}%
             \else \immediate\write\rw@write{\the\rw@toks}%
             \let\n@xt=\rw@begin\fi\fi \fi}
\def\rw@next{\rwr@teswitch\n@xt}
\def\rw@@d{\backtotext} \let\rw@end=\relax
\let\backtotext=\relax

\newdimen\refindent     \refindent=30pt
\def\refitem#1{\par \hangafter=0 \hangindent=\refindent \Textindent{#1}}
\def\REFNUM#1{\space@ver{}\refch@ck \firstreflinetrue%
 \global\advance\referencecount by 1 \xdef#1{\the\referencecount}}
\def\refnum#1{\space@ver{}\refch@ck \firstreflinetrue%
 \global\advance\referencecount by 1 \xdef#1{\the\referencecount}\refend}

\def\REF#1{\REFNUM#1%
 \immediate\write\referencewrite{%
 \noexpand\refitem{#1.}}%
\begingroup\obeyendofline\rw@start}
\def\ref{\refnum\?%
 \immediate\write\referencewrite{\noexpand\refitem{\?.}}%
\begingroup\obeyendofline\rw@start}
\def\Ref#1{\refnum#1%
 \immediate\write\referencewrite{\noexpand\refitem{#1.}}%
\begingroup\obeyendofline\rw@start}
\def\REFS#1{\REFNUM#1\global\lastrefsbegincount=\referencecount
\immediate\write\referencewrite{\noexpand\refitem{#1.}}%
\begingroup\obeyendofline\rw@start}
\newcount\figurecount     \figurecount=0
\newif\iffigureopen       \newwrite\figurewrite
\def\figch@ck{\chardef\rw@write=\figurewrite \iffigureopen\else
   \immediate\openout\figurewrite=figures.texauxil
   \figureopentrue\fi}
\def\FIGNUM#1{\space@ver{}\figch@ck \firstreflinetrue%
 \global\advance\figurecount by 1 \xdef#1{\the\figurecount}}
\def\FIG#1{\FIGNUM#1
   \immediate\write\figurewrite{\noexpand\refitem{#1.}}%
   \begingroup\obeyendofline\rw@start}
\def\fig{\FIGNUM\? fig.~\?%
\immediate\write\figurewrite{\noexpand\refitem{\?.}}%
\begingroup\obeyendofline\rw@start}
\def\figure{\FIGNUM\? figure~\?
   \immediate\write\figurewrite{\noexpand\refitem{\?.}}%
   \begingroup\obeyendofline\rw@start}
\def\Fig{\FIGNUM\? Fig.~\?%
\immediate\write\figurewrite{\noexpand\refitem{\?.}}%
\begingroup\obeyendofline\rw@start}
\def\Figure{\FIGNUM\? Figure~\?%
\immediate\write\figurewrite{\noexpand\refitem{\?.}}%
\begingroup\obeyendofline\rw@start}
\newcount\tablecount     \tablecount=0
\newif\iftableopen       \newwrite\tablewrite
\def\tabch@ck{\chardef\rw@write=\tablewrite \iftableopen\else
   \immediate\openout\tablewrite=tables.texauxil
   \tableopentrue\fi}
\def\TABNUM#1{\space@ver{}\tabch@ck \firstreflinetrue%
 \global\advance\tablecount by 1 \xdef#1{\the\tablecount}}
\def\TABLE#1{\TABNUM#1
   \immediate\write\tablewrite{\noexpand\refitem{#1.}}%
   \begingroup\obeyendofline\rw@start}
\def\Table{\TABNUM\? Table~\?%
\immediate\write\tablewrite{\noexpand\refitem{\?.}}%
\begingroup\obeyendofline\rw@start}
%

%
%
%
\def\masterreset{\global\pagenumber=1 \global\chapternumber=0
   \ifnum\the\equanumber<0\else \global\equanumber=0\fi
   \global\sectionnumber=0
   \global\referencecount=0 \global\figurecount=0 \global\tablecount=0 }
\def\FRONTPAGE{\ifvoid255\else\vfill\penalty-2000\fi
      \masterreset\global\frontpagetrue
      \global\lastf@@t=0 \global\footsymbolcount=0}

\def\paperstyle{\letterstylefalse\normalspace\papersize}
\def\letterstyle{\letterstyletrue\singlespace\lettersize}
\def\papersize{\hsize=6.5truein\vsize=9.1truein\hoffset=-.3truein
               \voffset=-.4truein\skip\footins=\bigskipamount}
\def\lettersize{\hsize=6.5truein\vsize=9.1truein\hoffset=-.3truein
    \voffset=.1truein\skip\footins=\smallskipamount \multiply
    \skip\footins by 3 }
\paperstyle   
%
%
\def\MEMO{\letterstyle\FRONTPAGE \letterfrontheadline={\hfil}
    \line{\quad\fourteenrm CERN MEMORANDUM\hfil\twelverm\the\date\quad}
    \medskip \memod@f}

\def\memit@m#1{\smallskip \hangafter=0 \hangindent=1in
      \Textindent{\caps #1}}
\def\memod@f{\xdef\mto{\memit@m{To:}}\xdef\from{\memit@m{From:}}%
     \xdef\topic{\memit@m{Topic:}}\xdef\subject{\memit@m{Subject:}}%
     \xdef\rule{\bigskip\hrule height 1pt\bigskip}}
\memod@f
\newskip\lettertopfil
\lettertopfil = 0pt plus 1.5in minus 0pt
\newskip\letterbottomfil
\letterbottomfil = 0pt plus 2.3in minus 0pt
\newskip\spskip \setbox0\hbox{\ } \spskip=-1\wd0
\def\addressee#1{\medskip\rightline{\the\date\hskip 30pt} \bigskip
   \vskip\lettertopfil
   \ialign to\hsize{\strut ##\hfil\tabskip 0pt plus \hsize \cr #1\crcr}
   \medskip\noindent\hskip\spskip}
\newskip\signatureskip       \signatureskip=40pt
\def\signed#1{\par \penalty 9000 \bigskip \dt@pfalse
  \everycr={\noalign{\ifdt@p\vskip\signatureskip\global\dt@pfalse\fi}}
  \setbox0=\vbox{\singlespace \halign{\tabskip 0pt \strut ##\hfil\cr
   \noalign{\global\dt@ptrue}#1\crcr}}
  \line{\hskip 0.5\hsize minus 0.5\hsize \box0\hfil} \medskip }

\def\endletter{\ifnum\pagenumber=1 \vskip\letterbottomfil\supereject
\else \vfil\supereject \fi}
\newbox\letterb@x
\def\lettertext{\par\unvcopy\letterb@x\par}
\def\multiletter{\setbox\letterb@x=\vbox\bgroup
      \everypar{\vrule height 1\baselineskip depth 0pt width 0pt }
      \singlespace \topskip=\baselineskip }
\def\letterend{\par\egroup}
%
%
%
\newskip\frontpageskip
\newtoks\pubtype
\newtoks\Pubnum
\newtoks\pubnum
\newtoks\pubnu
\newtoks\pubn
\newif\ifp@bblock  \p@bblocktrue
\def\PH@SR@V{\doubl@true \baselineskip=24.1pt plus 0.2pt minus 0.1pt
             \parskip= 3pt plus 2pt minus 1pt }
\def\PHYSREV{\paperstyle\PhysRevtrue\PH@SR@V}
\def\titlepage{\FRONTPAGE\paperstyle\ifPhysRev\PH@SR@V\fi
   \ifp@bblock\p@bblock\fi}
\def\nopubblock{\p@bblockfalse}
\def\endpage{\vfil\break}
\frontpageskip=1\medskipamount plus .5fil
\pubtype={\tensl Preliminary Version}
\Pubnum={$\rm CERN-TH.\the\pubnum $}
\pubnum={0000}
\def\p@bblock{\begingroup \tabskip=\hsize minus \hsize
   \baselineskip=1.5\ht\strutbox \topspace-2\baselineskip
   \halign to\hsize{\strut ##\hfil\tabskip=0pt\crcr
   \the \pubn\cr
   \the \Pubnum\cr
   \the \pubnu\cr
   \the \date\cr}\endgroup}
\def\title#1{\vskip\frontpageskip \titlestyle{#1} \vskip\headskip }
\def\author#1{\vskip\frontpageskip\titlestyle{\twelvecp #1}\nobreak}

\def\address#1{\par\kern 5pt\titlestyle{\twelvepoint\it #1}}
\def\andaddress{\par\kern 5pt \centerline{\sl and} \address}

\def\abstract{\vskip\frontpageskip\centerline{\fourteenrm ABSTRACT}
              \vskip\headskip }

%
%
%

\def\\{\relax\ifmmode\backslash\else$\backslash$\fi}
\def\globaleqnumbers{\relax\ifnum\the\equanumber<0%
\else\global\equanumber=-1\fi}
\def\nextline{\unskip\nobreak\hskip\parfillskip\break}

\def\journal#1&#2(#3){\unskip, \sl #1~\bf #2 \rm (19#3) }

\def\topspace{\hrule height 0pt depth 0pt \vskip}

\let\int=\intop         
\def\prop{\mathrel{{\mathchoice{\pr@p\scriptstyle}{\pr@p\scriptstyle}{
                \pr@p\scriptscriptstyle}{\pr@p\scriptscriptstyle} }}}
\def\pr@p#1{\setbox0=\hbox{$\cal #1 \char'103$}
   \hbox{$\cal #1 \char'117$\kern-.4\wd0\box0}}
\def\lsim{\mathrel{\mathpalette\@versim<}}
\def\gsim{\mathrel{\mathpalette\@versim>}}
\def\@versim#1#2{\lower0.2ex\vbox{\baselineskip\z@skip\lineskip\z@skip
  \lineskiplimit\z@\ialign{$\m@th#1\hfil##\hfil$\crcr#2\crcr\sim\crcr}}}
\def\leftrightarrowfill{$\m@th \mathord- \mkern-6mu
        \cleaders\hbox{$\mkern-2mu \mathord- \mkern-2mu$}\hfil
        \mkern-6mu \mathord\leftrightarrow$}
\def\lrover#1{\vbox{\ialign{##\crcr
        \leftrightarrowfill\crcr\noalign{\kern-1pt\nointerlineskip}
        $\hfil\displaystyle{#1}\hfil$\crcr}}}
%
%
%
\let\sec@nt=\sec
\def\sec{\relax\ifmmode\let\n@xt=\sec@nt\else\let\n@xt\section\fi\n@xt}
\def\obsolete#1{\message{Macro \string #1 is obsolete.}}
\def\firstsec#1{\obsolete\firstsec \section{#1}}
\def\firstsubsec#1{\obsolete\firstsubsec \subsection{#1}}
\def\thispage#1{\obsolete\thispage \global\pagenumber=#1\frontpagefalse}
\def\thischapter#1{\obsolete\thischapter \global\chapternumber=#1}
\def\nextequation#1{\obsolete\nextequation \global\equanumber=#1
   \ifnum\the\equanumber>0 \global\advance\equanumber by 1 \fi}
\def\BOXITEM{\afterassigment\B@XITEM\setbox0=}
\def\B@XITEM{\par\hangindent\wd0 \noindent\box0 }
%

%
%

%
%

%
%

%

%

%

%

%

%
%
%
\def\boxit#1{\vbox{\hrule\hbox{\vrule\kern3pt\vbox{\kern3pt#1\kern3pt}
\kern3pt\vrule}\hrule}}
%
%
%
\catcode`@=12 
\message{ by V.K./U.B.}
\everyjob{\input imyphyx }
%
%
%
%
%
%
%
%
%
%
%
%
%
\catcode`@=11

\font\seventeencp=cmcsc10 scaled\magstep3
\def\SIZE{\hsize=6.6truein\vsize=9.1truein}
\def\OFFSET{\voffset=1.2truein\hoffset=.8truein}
\def\papersize{\SIZE\OFFSET\skip\footins=\bigskipamount
\normaldisplayskip= 30pt plus 5pt minus 10pt}
\Pubnum={\rm CERN$-$TH.\the\pubnum }
\def\title#1{\vskip\frontpageskip\vskip .50truein
     \titlestyle{\seventeencp #1} \vskip\headskip\vskip\frontpageskip
     \vskip .2truein}
\def\author#1{\vskip .27truein\titlestyle{#1}\nobreak}

\def\p@bblock{\begingroup \tabskip=\hsize minus \hsize
   \baselineskip=1.5\ht\strutbox \topspace-2\baselineskip
   \halign to\hsize{\strut ##\hfil\tabskip=0pt\crcr
   \the \Pubnum\cr}\endgroup}
\def\makefootline{\iffrontpage\vskip .27truein\line{\the\footline}
                 \vskip -.1truein\line{\the\date\hfil}
              \else\line{\the\footline}\fi}
\paperfootline={\iffrontpage \the\Pubnum\hfil\else\hfil\fi}
\paperheadline={\iffrontpage\hfil
                \else\twelverm\hss $-$\ \folio\ $-$\hss\fi}
\newif\ifmref  
\newif\iffref  
\def\xrefsend{\xrefmark{\count255=\referencecount
\advance\count255 by-\lastrefsbegincount
\ifcase\count255 \number\referencecount
\or \number\lastrefsbegincount,\number\referencecount
\else \number\lastrefsbegincount-\number\referencecount \fi}}
\def\xrefsdub{\xrefmark{\count255=\referencecount
\advance\count255 by-\lastrefsbegincount
\ifcase\count255 \number\referencecount
\or \number\lastrefsbegincount,\number\referencecount
\else \number\lastrefsbegincount,\number\referencecount \fi}}
\def\xREFNUM#1{\space@ver{}\refch@ck\firstreflinetrue%
\global\advance\referencecount by 1
\xdef#1{\xrefend}}
\def\xrefend{\xrefmark{\number\referencecount}}
\def\xrefmark#1{[{#1}]}
\def\xRef#1{\xREFNUM#1\immediate\write\referencewrite%
{\noexpand\refitem{#1}}\begingroup\obeyendofline\rw@start}%
\def\xREFS#1{\xREFNUM#1\global\lastrefsbegincount=\referencecount%
\immediate\write\referencewrite{\noexpand\refitem{#1}}%
\begingroup\obeyendofline\rw@start}
\def\rrr#1#2{\relax\ifmref{\iffref\xREFS#1{#2}%
\else\xRef#1{#2}\fi}\else\xRef#1{#2}\xrefend\fi}
\referencecount=0
%
\space@ver{}\refch@ck\firstreflinetrue%
\immediate\write\referencewrite{}%
\begingroup\obeyendofline\rw@start{}%
\def\plb#1({Phys.\ Lett.\ $\underline  {#1B}$\ (}
\def\nup#1({Nucl.\ Phys.\ $\underline {B#1}$\ (}
\def\plt#1({Phys.\ Lett.\ $\underline  {B#1}$\ (}
\def\cmp#1({Comm.\ Math.\ Phys.\ $\underline  {#1}$\ (}
\def\prp#1({Phys.\ Rep.\ $\underline  {#1}$\ (}
\def\prl#1({Phys.\ Rev.\ Lett.\ $\underline  {#1}$\ (}
\def\prv#1({Phys.\ Rev. $\underline  {D#1}$\ (}
\def\und#1({            $\underline  {#1}$\ (}
\message{ by W.L.}
\everyjob{\input offset }
\catcode`@=12

\let\it=\sl

%

%
\newbox\hdbox%
\newcount\hdrows%
\newcount\multispancount%
\newcount\ncase%
\newcount\ncols
\newcount\nrows%
\newcount\nspan%
\newcount\ntemp%
\newdimen\hdsize%
\newdimen\newhdsize%
\newdimen\parasize%
\newdimen\spreadwidth%
\newdimen\thicksize%
\newdimen\thinsize%
\newdimen\tablewidth%
\newif\ifcentertables%
\newif\ifendsize%
\newif\iffirstrow%
\newif\iftableinfo%
\newtoks\dbt%
\newtoks\hdtks%
\newtoks\savetks%
\newtoks\tableLETtokens%
\newtoks\tabletokens%
\newtoks\widthspec%
%
%
\immediate\write15{%
CP SMSG GJMSINK TEXTABLE --> TABLE MACROS V. 851121 JOB = \jobname%
}%
%
%
\tableinfotrue%
\catcode`\@=11
%
%
\def\tstrut{\vrule height3.1ex depth1.2ex width0pt}%
\def\and{\char`\&}
\def\tablerule{\noalign{\hrule height\thinsize depth0pt}}%
\thicksize=1.5pt
\thinsize=0.6pt
\def\thickrule{\noalign{\hrule height\thicksize depth0pt}}%
\def\ctr#1{\hfil\ #1\hfil}%
%
%
%
%
\tablewidth=-\maxdimen%
\spreadwidth=-\maxdimen%
\def\tabskipglue{0pt plus 1fil minus 1fil}%
%
%
\centertablestrue%
%
%
%
%
\parasize=4in%
\gdef\ARGS{########}
\gdef\headerARGS{####}
\def\@mpersand{&}
{\catcode`\|=13
\gdef\letbarzero{\let|0}
\gdef\letbartab{\def|{&&}}%
\gdef\letvbbar{\let\vb|}%
}
{\catcode`\&=4
\def\ampskip{&\omit\hfil&}
\catcode`\&=13
\let&0
\xdef\letampskip{\def&{\ampskip}}%
\gdef\letnovbamp{\let\novb&\let\tab&}
}
\def\begintable{
   \begingroup%
   \catcode`\|=13\letbartab\letvbbar%
   \catcode`\&=13\letampskip\letnovbamp%
   \def\multispan##1{
      \omit \mscount##1%
      \multiply\mscount\tw@\advance\mscount\m@ne%
      \loop\ifnum\mscount>\@ne \sp@n\repeat%
   }
   \def\|{%
      &\omit\widevline&%
   }%
   \ruledtable
}
\long\def\ruledtable#1\endtable{%
%
%
%
   \offinterlineskip
   \tabskip 0pt
   \def\widevline{\vrule width\thicksize}
   \def\endrow{\@mpersand\omit\hfil\crnorm\@mpersand}%
   \def\crthick{\@mpersand\crnorm\thickrule\@mpersand}%
   \def\crthickneg##1{\@mpersand\crnorm\thickrule
          \noalign{{\skip0=##1\vskip-\skip0}}\@mpersand}%
   \def\crnorule{\@mpersand\crnorm\@mpersand}%
   \def\crnoruleneg##1{\@mpersand\crnorm
          \noalign{{\skip0=##1\vskip-\skip0}}\@mpersand}%
   \let\nr=\crnorule
   \def\endtable{\@mpersand\crnorm\thickrule}%
   \let\crnorm=\cr
%
%
   \edef\cr{\@mpersand\crnorm\tablerule\@mpersand}%
   \def\crneg##1{\@mpersand\crnorm\tablerule
          \noalign{{\skip0=##1\vskip-\skip0}}\@mpersand}%
   \let\ctneg=\crthickneg
   \let\nrneg=\crnoruleneg
   \the\tableLETtokens
%
%
   \tabletokens={&#1}
%
%
   \countROWS\tabletokens\into\nrows%
   \countCOLS\tabletokens\into\ncols%
%
%
   \advance\ncols by -1%
   \divide\ncols by 2%
   \advance\nrows by 1%
%
%
   \iftableinfo %
      \immediate\write16{[Nrows=\the\nrows, Ncols=\the\ncols]}%
   \fi%
%
%
   \ifcentertables
      \ifhmode \par\fi
      \line{
      \hss
   \else %
      \hbox{%
   \fi
      \vbox{%
         \makePREAMBLE{\the\ncols}
         \edef\next{\preamble}
         \let\preamble=\next
         \makeTABLE{\preamble}{\tabletokens}
      }
      \ifcentertables \hss}\else }\fi
   \endgroup
   \tablewidth=-\maxdimen
   \spreadwidth=-\maxdimen
}
\def\makeTABLE#1#2{
   {
   \let\ifmath0
   \let\header0
   \let\multispan0
%
%
   \ncase=0%
   \ifdim\tablewidth>-\maxdimen \ncase=1\fi%
   \ifdim\spreadwidth>-\maxdimen \ncase=2\fi%
   \relax
%
   \ifcase\ncase %
      \widthspec={}%
   \or %
      \widthspec=\expandafter{\expandafter t\expandafter o%
                 \the\tablewidth}%
   \else %
      \widthspec=\expandafter{\expandafter s\expandafter p\expandafter r%
                 \expandafter e\expandafter a\expandafter d%
                 \the\spreadwidth}%
   \fi %
   \xdef\next{
      \halign\the\widthspec{%
      #1
      \noalign{\hrule height\thicksize depth0pt}
      \the#2\endtable
%
      }
   }
   }
   \next
}
\def\makePREAMBLE#1{
   \ncols=#1
   \begingroup
   \let\ARGS=0
   \edef\xtp{\widevline\ARGS\tabskip\tabskipglue%
   &\ctr{\ARGS}\tstrut}
   \advance\ncols by -1
   \loop
      \ifnum\ncols>0 %
      \advance\ncols by -1%
      \edef\xtp{\xtp&\vrule width\thinsize\ARGS&\ctr{\ARGS}}%
   \repeat
   \xdef\preamble{\xtp&\widevline\ARGS\tabskip0pt%
   \crnorm}
   \endgroup
}
\def\countROWS#1\into#2{
   \let\countREGISTER=#2%
   \countREGISTER=0%
   \expandafter\ROWcount\the#1\endcount%
}%
\def\ROWcount{%
   \afterassignment\subROWcount\let\next= %
}%
\def\subROWcount{%
   \ifx\next\endcount %
      \let\next=\relax%
   \else%
      \ncase=0%
      \ifx\next\cr %
         \global\advance\countREGISTER by 1%
         \ncase=0%
      \fi%
      \ifx\next\endrow %
         \global\advance\countREGISTER by 1%
         \ncase=0%
      \fi%
      \ifx\next\crthick %
         \global\advance\countREGISTER by 1%
         \ncase=0%
      \fi%
      \ifx\next\crnorule %
         \global\advance\countREGISTER by 1%
         \ncase=0%
      \fi%
      \ifx\next\crthickneg %
         \global\advance\countREGISTER by 1%
         \ncase=0%
      \fi%
      \ifx\next\crnoruleneg %
         \global\advance\countREGISTER by 1%
         \ncase=0%
      \fi%
      \ifx\next\crneg %
         \global\advance\countREGISTER by 1%
         \ncase=0%
      \fi%
      \ifx\next\header %
         \ncase=1%
      \fi%
      \relax%
      \ifcase\ncase %
         \let\next\ROWcount%
      \or %
         \let\next\argROWskip%
      \else %
      \fi%
   \fi%
   \next%
}
\def\counthdROWS#1\into#2{%
\dvr{10}%
   \let\countREGISTER=#2%
   \countREGISTER=0%
\dvr{11}%
\dvr{13}%
   \expandafter\hdROWcount\the#1\endcount%
\dvr{12}%
}%
\def\hdROWcount{%
   \afterassignment\subhdROWcount\let\next= %
}%
\def\subhdROWcount{%
   \ifx\next\endcount %
      \let\next=\relax%
   \else%
      \ncase=0%
      \ifx\next\cr %
         \global\advance\countREGISTER by 1%
         \ncase=0%
      \fi%
      \ifx\next\endrow %
         \global\advance\countREGISTER by 1%
         \ncase=0%
      \fi%
      \ifx\next\crthick %
         \global\advance\countREGISTER by 1%
         \ncase=0%
      \fi%
      \ifx\next\crnorule %
         \global\advance\countREGISTER by 1%
         \ncase=0%
      \fi%
      \ifx\next\header %
         \ncase=1%
      \fi%
\relax%
      \ifcase\ncase %
         \let\next\hdROWcount%
      \or%
         \let\next\arghdROWskip%
      \else %
      \fi%
   \fi%
   \next%
}%
{\catcode`\|=13\letbartab
\gdef\countCOLS#1\into#2{%
   \let\countREGISTER=#2%
   \global\countREGISTER=0%
   \global\multispancount=0%
   \global\firstrowtrue
   \expandafter\COLcount\the#1\endcount%
   \global\advance\countREGISTER by 3%
   \global\advance\countREGISTER by -\multispancount
}%
\gdef\COLcount{%
   \afterassignment\subCOLcount\let\next= %
}%
{\catcode`\&=13%
\gdef\subCOLcount{%
   \ifx\next\endcount %
      \let\next=\relax%
   \else%
      \ncase=0%
      \iffirstrow
         \ifx\next& %
            \global\advance\countREGISTER by 2%
            \ncase=0%
         \fi%
         \ifx\next\span %
            \global\advance\countREGISTER by 1%
            \ncase=0%
         \fi%
         \ifx\next| %
            \global\advance\countREGISTER by 2%
            \ncase=0%
         \fi
         \ifx\next\|
            \global\advance\countREGISTER by 2%
            \ncase=0%
         \fi
         \ifx\next\multispan
            \ncase=1%
            \global\advance\multispancount by 1%
         \fi
         \ifx\next\header
            \ncase=2%
         \fi
         \ifx\next\cr       \global\firstrowfalse \fi
         \ifx\next\endrow   \global\firstrowfalse \fi
         \ifx\next\crthick  \global\firstrowfalse \fi
         \ifx\next\crnorule \global\firstrowfalse \fi
         \ifx\next\crnoruleneg \global\firstrowfalse \fi
         \ifx\next\crthickneg  \global\firstrowfalse \fi
         \ifx\next\crneg       \global\firstrowfalse \fi
      \fi
\relax
      \ifcase\ncase %
         \let\next\COLcount%
      \or %
         \let\next\spancount%
      \or %
         \let\next\argCOLskip%
      \else %
      \fi %
   \fi%
   \next%
}%
\gdef\argROWskip#1{%
   \let\next\ROWcount \next%
}
\gdef\arghdROWskip#1{%
   \let\next\ROWcount \next%
}
\gdef\argCOLskip#1{%
   \let\next\COLcount \next%
}
}
}
\def\spancount#1{
   \nspan=#1\multiply\nspan by 2\advance\nspan by -1%
   \global\advance \countREGISTER by \nspan
   \let\next\COLcount \next}%
\def\dvr#1{\relax}%
\def\header#1{%
\dvr{1}{\let\cr=\@mpersand%
\hdtks={#1}%
\counthdROWS\hdtks\into\hdrows%
\advance\hdrows by 1%
\ifnum\hdrows=0 \hdrows=1 \fi%
\dvr{5}\makehdPREAMBLE{\the\hdrows}%
\dvr{6}\getHDdimen{#1}%
{\parindent=0pt\hsize=\hdsize{\let\ifmath0%
\xdef\next{\valign{\headerpreamble #1\crnorm}}}\dvr{7}\next\dvr{8}%
}%
}\dvr{2}}
\def\makehdPREAMBLE#1{
\dvr{3}%
\hdrows=#1
{
\let\headerARGS=0%
\let\cr=\crnorm%
\edef\xtp{\vfil\hfil\hbox{\headerARGS}\hfil\vfil}%
\advance\hdrows by -1
\loop
\ifnum\hdrows>0%
\advance\hdrows by -1%
\edef\xtp{\xtp&\vfil\hfil\hbox{\headerARGS}\hfil\vfil}%
\repeat%
\xdef\headerpreamble{\xtp\crcr}%
}
\dvr{4}}
\def\getHDdimen#1{%
\hdsize=0pt%
\getsize#1\cr\end\cr%
}
\def\getsize#1\cr{%
\endsizefalse\savetks={#1}%
\expandafter\lookend\the\savetks\cr%
\relax \ifendsize \let\next\relax \else%
\setbox\hdbox=\hbox{#1}\newhdsize=1.0\wd\hdbox%
\ifdim\newhdsize>\hdsize \hdsize=\newhdsize \fi%
\let\next\getsize \fi%
\next%
}%
\def\lookend{\afterassignment\sublookend\let\looknext= }%
\def\sublookend{\relax%
\ifx\looknext\cr %
\let\looknext\relax \else %
   \relax
   \ifx\looknext\end \global\endsizetrue \fi%
   \let\looknext=\lookend%
    \fi \looknext%
}%
%
%
\def\tablelet#1{%
   \tableLETtokens=\expandafter{\the\tableLETtokens #1}%
}%
\catcode`\@=12
\def\CLMR{\rrr\CLMR{J.A. Casas, Z. Lalak, C. Mu\~noz and
G.G. Ross, \nup347 (1990) 243;\nextline
L. Dixon, SLAC preprint 5229 (1990).}}

\def\OW{\rrr\OW{B. Ovrut and J. Wess, \plb119 (1982) 105.}}

\def\DFKZ{\rrr\DFKZ{J.P. Derendinger, S. Ferrara, C. Kounnas and
F. Zwirner,{\it ``On loop corrections to string effective field theories:
         field-dependent gauge couplings and sigma-model anomalies'',}
        preprint CERN-TH.6004/91, LPTENS 91-4                   (1991).}}

\def\AELN{\rrr\AELN{I. Antoniadis, J. Ellis, R. Lacaze and D.V.
Nanopoulos, \plb268 (1991) 188;     S. Kalara, J.L. L\'opez and
D.V. Nanopoulos, \plb269 (1991) 84;      J. Ellis, S.Kelley and
D.V. Nanopoulos, CERN-TH.6140/91 (1991).}}

\def\FAY{\rrr\FAY{P. Fayet, \plb69 (1977) 489;\plb84 (1979) 416;
\plb78 (1978)  417;\nextline
P. Fayet and G. Farrar, \plb79 (1978) 442;\plb89 (1980) 191.}}

\def\DFS{\rrr\DFS{M. Dine, W. Fischler and M. Srednicki,
\nup189 (1981) 575;\nup202 (1982) 238;\nextline
S. Dimopoulos and S. Raby, \nup192 (1981) 353.}}

\def\MARTI{\rrr\MARTI{G. Martinelli, Review talk at the Symposium on
Lepton-Photon Interactions, Geneva 1991.}}

\def\ABJ{\rrr\ABJ{S. Adler, Phys.Rev. 177 (1969) 2426; \nextline
J.S. Bell and R. Jackiw, Nuovo Cimento 60A (1969) 47.}}

\def\WK{\rrr\WK{L. Krauss and F. Wilczek, Phys.Rev.Lett. 62 (1989) 1221.}}

\def\DIS{\rrr\DIS{T. Banks, \nup323 (1989) 90; \nextline
L. Krauss, Gen.Rel.Grav. 22 (1990) 50; \nextline
M. Alford, J. March-Russell and F. Wilczek, \nup337 (1990) 695;\nextline
J. Preskill and L. Krauss, \nup341 (1990) 50;\nextline
M. Alford, S. Coleman and J. March-Russell, preprint HUTP-90/A040 (1990).
}}

\def\WH{\rrr\WH{For a review and references see: T. Banks, Santa Cruz preprint
SCIPP 89/17 (1989).}}

\def\PR{\rrr\PR{J. Preskill, preprint CALT-68-1493 (1990).}}

\def\GRA{\rrr\GRA{R. Delbourgo and A. Salam, \plb40 (1972) 381;\nextline
T. Eguchi and P. Freund, Phys.Rev.Lett. 37 (1976) 1251; \nextline
L. Alvarez-Gaum\'e and E. Witten, \nup234 (1983) 269. }}

\def\IR{\rrr\IR{L.E. Ib\'a\~nez and G.G. Ross, in preparation.}}

\def\BHR{\rrr\BHR{L. Hall and M. Suzuki, \nup231 (1984) 419;\nextline
M. Bento, L. Hall and G.G. Ross, \nup292 (1987) 400.}}

\def\ZDOS{\rrr\ZDOS{A. Font, L.E. Ib\'a\~nez and F. Quevedo, \plb2288 (1989) 79
; \nextline
A. Font, L.E. Ib\'a\~nez and  F. Quevedo  \nup345 (1990) 389.}}

\def\BRAN{\rrr\BRAN{ L. Perivolaropoulos, A. Matheson, A.C. Davis and
R. Brandenberger, preprint BROWN-HET-739 (1990).}}

\def\VIRGIN{\rrr\VIRGIN{ L.E. Ib\'a\~nez, Proceedings of the 5-th ASI on
Techniques and Concepts of High Energy Physics, St. Croix (Virgin Islands),
July 14-25, 1988. Edited by T. Ferbel, Plenum Press (1989); \nextline
A. Font, L.E. Ib\'a\~nez and F. Quevedo, \plb228 (1989) 79;\nextline
C. Geng and R. Marshak, Phys.Rev. D39 (1989) 693; \nextline
J. Minahan, P. Ramond and R. Warner, Phys.Rev. D41 (1990) 715.}}
\def\SS{\rrr\SS{H.P. Nilles, Phys.Rep. C110 (1984) 1;\nextline
G.G. Ross, ``Grand Unified Theories'', Benjamin Inc., (1984);\nextline
L.E. Ib\`a\~nez, ``Beyond the Standard Model (yet again)'',
CERN-TH.5982/91, to appear in the proceedings of the 1990 CERN
summer school (Mallorca).}}

\def\RPB{\rrr\RPB{L. Hall and M. Suzuki, \nup231        (1984) 419;
\nextline  F. Zwirner, \plb132 (1983) 103;\nextline
R. Mohapatra, Phys.Rev.D34 (1986) 3457;\nextline
R. Barbieri and A. Masiero, \nup267 (1986) 679;\nextline
S. Dimopoulos and L. Hall, \plb196 (1987) 135\nextline
V. Barger, G.F. Giudice and T. Han, Phys.Rev.D40  (1989) 2987.}}

\def\RP{\rrr\RP{S. Dimopoulos, R. Esmaizadeh, L. Hall and G. Starkman,
Phys.Rev.D41 (1990) 2099;\nextline
S. Dawson, \nup261 (1985) 297. }}

\def\DR{\rrr\DR{
H. Dreiner and G.G. Ross, Oxford preprint OUTP-91-15P\nextline
P. Binetruy et al., Proceedings of the ECFA Large Hadron Collider (LHC)
Workshop, Aachen, 1990. Vol.I, CERN report CERN 90-10.}}

\def\DRB{\rrr\DRB{H. Dreiner and G.G. Ross, Oxford preprint, in
preparation.}}

\def\WHH{\rrr\WHH{G. Gilbert, \nup328 (1989) 159  and references therein.
}}

\def\IR{\rrr\IR{ L.E. Ib\`a\~nez and G.G. Ross, CERN-TH-6000/91 (1991),
to appear in Phys.Lett.B.}}

\def\EEE{\rrr\EEE{J. Ellis, J. Hagelin, D. Nanopoulos and K. Tamvakis,
\plb124 (1983) 484.}}

\def\KAP{\rrr\KAP{V. Kaplunovsky, Phys.Rev.Lett. 55 (1985) 1036;
\nextline M. Dine and N. Seiberg, \plb162 (1985) 299.}}

\def\CAMP{\rrr\CAMP{B. Campbell, S. Davidson, J. Ellis and K. Olive,
\plb256 (1991) 457 ;\nextline
W. Fischler, G. Giudice, R. Leigh and S. Paban,
\plb258 (1991) 45;\nextline
H. Dreiner and G.G. Ross, in preparation. }}

\def\YAN{\rrr\YAN{M. Fukugita and T. Yanagida, \plb174 (1986) 45;
Phys.Rev.D42 (1990) 1285;
\nextline A. Bouquet and P. Salati, \nup284 (1987) 557;
\nextline  J. Harvey and M. Turner, Phys.Rev.D42 (1990) 3344;\nextline
A. Nelson and S. Barr, \plb246 (1990) 141. }}

\def\DINE{\rrr\DINE{I. Affleck and M. Dine, \nup249 (1985) 361;\nextline
D. Morgan, Texas preprint UTTG-04-91 (1991).}}

\def\HALLL{\rrr\HALLL{S. Dimopoulos and L. Hall, \plb196 (1987) 135;
\nextline J. Cline and S. Raby, Ohio preprint DOE/ER/01545-444 (1990).}}

\def\BGH{\rrr\BGH{R. Barbieri, M. Guzzo, A. Masiero and D. Tommasini,
\plb252 (1990) 251.}}

\def\PET{\rrr\PET{M. Guzzo, A. Masiero and S. Petcov,
SISSA preprint 16/91 EP (1991);   \nextline
E. Roulet, Fermilab preprint 91/18-A (1991) .}}

\def\FIQQ{\rrr\FIQQ{A. Font, L.E. Ib\`a\~nez and F. Quevedo,
\plb228 (1989) 79.}}

\def\UGO{\rrr\UGO{
J. Ellis, S. Kelley and D.V. Nanopoulos, \plb249 (1990) 441;
\plb260 (1991) 131;\nextline
P. Langacker           , "Precision tests of the standard model",
Pennsylvania preprint UPR-0435T, (1990);\nextline
U. Amaldi, W. de Boer and H. F\"urstenau, CERN-PPE/91-44 (1991);\nextline
P. Langacker and M. Luo, Pennsylvania preprint UPR-0466T, (1991).}}

\def\KAC{\rrr\KAC{ A. Font, L.E. Ib\`a\~nez and F. Quevedo,
\nup345 (1990) 389;\nextline
J. Ellis, J. L\'opez and D.V. Nanopoulos,  \plb245 (1990) 375.}}

\def\DG{\rrr\DG{S. Dimopoulos and H. Georgi, \nup193 (1981) 150.}}

\def\CHSW{\rrr\CHSW{P. Candelas, G. Horowitz,
A. Strominger and E. Witten, \nup 258 (1985) 46.}}
\def\DUAL{\rrr\DUAL{
K. Kikkawa and M. Yamasaki, \plb149 (1984) 357;
B. Sathiapalan, \prl58 (1987) 1597;
V.P. Nair, A. Shapere, A. Strominger and F. Wilczek,
      \nup287 (1987) 402;
R.~Dijkgraaf, E.~Verlinde and H.~Verlinde, \cmp115 (1988) 649;
  preprint THU-87/30;
R. Brandenberger and C. Vafa, \nup316 (1989) 391;
A. Giveon, E. Rabinovici and G. Veneziano,
    \nup322 (1989) 167;
A. Shapere and F. Wilzcek, \nup320 (1989) 669.}}

\def\FLST{\rrr\FLST{S. Ferrara,
 D. L\"ust, A. Shapere and S. Theisen,
   \plt225 (1989) 363.}}

\def\WIT{\rrr\WIT{E. Witten, \plb155 (1985) 151. }}

\def\CFGVP{\rrr\CFGVP{E. Cremmer, S. Ferrara, L. Girardello and
A. Van Proeyen, \nup212 (1983) 413. }}
\def\SH{\rrr\SH{H. Georgi, \nup156 (1979) 126. }}
\def\ADBEL{\rrr\ADBEL{S. Adler, Phys.Rev. 177 (1969) 2426. \nextline
J.S. Bell and R. Jackiw, Nuovo Cimento 60A (1969) 47.}}
\def\SAL{\rrr\SAL{R. Delbourgo and A. Salam, \plb40 (1972) 381;
\nextline T. Eguchi and P. Freund, Phys.Rev.Lett. 37 (1976) 1251.}}
\def\ALWI{\rrr\ALWI{L. Alvarez-Gaum\'e and E. Witten, \nup234 (1983) 269.
}}
\def\ASI{\rrr\ASI{L. E. Ib\' a\~ nez, Proceedings of the 5-th ASI
on Techniques and Concepts of High Energy Physics, St. Croix (Virgin
Islands), July 14-25, 1988. Edited by T. Ferbel, Plenum Press (1989).
}}
\def\FIQUNO{\rrr\FIQUNO{A. Font, L.E. Ib\' a\~ nez and F. Quevedo,
Phys.Lett. B228 (1989) 79. }}
\def\MARS{\rrr\MARS{C. Geng and R. Marshak, Phys.Rev. D39 (1989) 693;
\nextline
J. Minahan, P. Ramond and R. Warner, Phys.Rev. D41 (1990) 715;\nextline
K. Babu and R. Mohapatra, Phys.Rev. D41 (1990) 271. }}
\def\PDG{\rrr\PDG{Particle Data Group, Phys.Lett. B204 (1988) 1. }}
\def\TOOFT{\rrr\TOOFT{G. 't Hooft, in "Recent Developments in Gauge
Theories", ed. by G. 't Hooft et al., Plenum Press, New York (1981).}}
\def\CP{\rrr\CP{ For reviews of the strong-CP problem see: \nextline
J. E. Kim, Phys.Rep. 150 (1987) 1; \nextline
H. Y. Cheng, Phys.Rep. 158 (1988) 1.}}
\def\ALT{\rrr\ALT{I. S. Altarev et al., JETP Lett. 44 (1986) 461.}}
\def\SMIT{\rrr\SMIT{K. M. Smith et al., Phys. Lett. B234 (1990) 191. }}
\def\PQ{\rrr\PQ{R. Peccei and H. Quinn, Phys.RevLett. 38 (1977) 1440;
Phys.Rev. D16 (1977) 1791. }}
\def\WEWI{\rrr\WEWI{S. Weinberg, Phys.Rev.Lett. 40 (1978) 223;\nextline
F. Wilczek, Phys.Rev.Lett. 40 (1978) 279.}}
\def\INVIS{\rrr\INVIS{J. E. Kim, Phys.Rev.Lett. 43 (1979) 103;\nextline
M. Shifman, A. Vainstein and V. Zakharov, \nup116 (1980) 493;\nextline
M. Dine, W. Fischler and M. Srednicki, \plb104 (1981) 99. }}
\def\DICUS{\rrr\DICUS{D. Dicus et al., \prv22 (1980) 839;\nextline
M. Fukugita, S. Watamura and M. Yoshimura, \prv26 (1982) 1840;\nextline
A. Pantziris and K. Kang, \prv33 (1986) 3509.}}
\def\SN{\rrr\SN{R. Mayle, J. Wilson, J. Ellis and K. Olive,
\plb219 (1989) 515.}}
\def\PRESK{\rrr\PRESK{J. Preskill, M. Wise and F. Wilczek, \plb120 (1983)
127;\nextline
L. Abbott and P. Sikivie, \plb120 (1983) 133;\nextline
M. Dine and W. Fischler, \plb120 (1983) 137.}}
\def\NELS{\rrr\NELS{A. Nelson, \plb136 (1984) 387; \plb143 (1984)
165;\nextline
S. Barr, Phys.Rev.Lett. 53 (1984) 329;\prv30 (1984) 1805;\nextline
S. Barr and A. Masiero, \prv38 (1988) 366.}}
\def\GILD{\rrr\GILD{E. Gildener, \prv14 (1976) 1667.}}
\def\COSM{\rrr\COSM{M. Veltman, Phys.Rev.Lett. 34 (1975) 77 ;\nextline
A. Linde, JETP.Lett. 19 (1974) 183. }}
\def\TECH{\rrr\TECH{S. Weinberg, \prv13 (1976) 974;\prv19 (1979) 1277;
\nextline
L. Susskind, \prv20 (1979) 2619;\nextline
E. Farhi and L. Susskind, Phys.Rep. 74C (1981) 2777.}}
\def\DIMSU{\rrr\DIMSU{S. Dimopoulos and L. Susskind, \nup155 (1979) 237;
E. Eichten and K. Lane, \plb90 (1980) 125.}}
\def\DIMEL{\rrr\DIMEL{S. Dimopoulos and J. Ellis, \nup182 (1981) 505.}}
\def\HOLDO{\rrr\HOLDO{B. Holdom, \prv24 (1981) 1441;\plb150 (1985) 301;
\nextline
K. Yamawaki, M. Bando and K. Matumoto, Phys.Rev.Lett. 56 (1986) 1335;
\nextline
T. Appelquist, D. Karabali and L. Wijewardhana, Phys.Rev.Lett. 57 (1986)
957;\nextline
T. Appelquist and L. Wijewardhana, \prv36 (1987) 568;\nextline
S. Raby and G. Giudice, Ohio preprint DOE/ER/01545-447 (1990).}}
\def\NAMB{\rrr\NAMB{Y. Nambu, in New Theories in Physics, proceedings
of the XIth Kazimierz Symposium, (1988). World Scientific (1989).}}
\def\LINDN{\rrr\LINDN{W. Bardeen, C. Hill and M. Lindner,
\prv41 (1990) 1647;\nextline
V. Miransky, M. Tanabashi and K. Yamawaki, Mod.Phys.Lett. A4 (1989) 1043;
\plb221 (1989) 177. }}
\def\CDFMT{\rrr\CDFMT{CDF Collaboration, Fermilab preprint Conf-90/138-E
\nextline (1990).}}
\def\FERN{\rrr\FERN{E. Fernandez, CERN-PPE/90-151, talk given at the
Neutrino-90 Conference, CERN, June 1990.}}
\def\NAJL{\rrr\NAJL{Y. Nambu and G. Jona-Lasinio, Phys. Rev. 122 (1961)
345.}}
\def\CDFLC{\rrr\CDFLC{CDF Collaboration (presented by S. Bertolucci),
Fermilab-Conf-90/45-T (1990). Proceedings of the 8th Topical
Workshop on $p-{\bar p}$ Collider Physics, Castiglione (1989).}}
\def\CHAGA{\rrr\CHAGA{M. Chanowitz and M. K. Gaillard, \nup261 (1985)
379;\nextline
M. Chanowitz, M. Golden and H. Georgi, \prv36 (1987) 149; Phys.Rev.Lett.
57 (1986) 2344.}}
\def\DOHE{\rrr\DOHE{A. Dobado and M. Herrero, \plb228 (1989) 495;
\plb233 (1989) 505;\nextline
J. F. Donaghue and C. Ramirez, \plb234 (1990) 361.}}
\def\ESPR{\rrr\ESPR{A. Dobado, D. Espriu and M. Herrero,
CERN-TH.5785/90 (1990).}}
\def\PT{\rrr\PT{M. Peskin and T. Takeuchi, SLAC-PUB-5272 (1990).}}
\def\RAND{\rrr\RAND{M. Golden and L. Randall, Fermilab preprint
FERMILAB-PUB-90/83-T (1990).}}
\def\MARCI{\rrr\MARCI{W. Marciano and J. Rosner, BNL-44997 (1990).}}
\def\RING{\rrr\RING{A. Ringwald, \nup330 (1990) 1.}}
\def\ESPI{\rrr\ESPI{O. Espinosa, Caltech preprint CALT-68-1586 (1989).}}
\def\TOOFTB{\rrr\TOOFTB{G. 't Hooft, Phys.Rev.Lett. 37 (1976) 8;
\prv14 (1976) 3422.}}
\def\RUBA{\rrr\RUBA{V. Kuzmin, V. Rubakov and M. Shaposnikov,
\plb155 (1985) 36.}}
\def\MANTO{\rrr\MANTO{F. Klinkhamer and N. Manton, \prv30 (1984) 2212.}}
\def\SUCIN{\rrr\SUCIN{H. Georgi and S. L. Glashow, Phys.Rev.Lett. 32
(1974) 32.}}
\def\PASA{\rrr\PASA{J. Pati and A. Salam, \prv10 (1974) 275.}}
\def\GQW{\rrr\GQW{H. Georgi, H. Quinn and S. Weinberg, Phys.Rev.Lett.
33 (1974) 451.}}
\def\GUTS{\rrr\GUTS{For a review on Grund Unified Theories see:
\nextline
G.G. Ross, "Grand Unified Theories", Benjamin Inc. (1984);
\nextline
P. Langacker, Phys.Rep. 72C (1981) 185.}}
\def\PDEC{\rrr\PDEC{IMB Collaboration, S. Seidel et al. Phys.Rev.Lett.
61 (1988) 2522;\nextline
Kamiokande-II Collaboration, K. Hirata et al. \plb220 (1989) 308;
\nextline
Ch. Berger et al., \nup313 (1989) 509;\nextline
T. Philips et al., \plb224 (1989) 348.}}
\def\MARSI{\rrr\MARSI{W. Marciano and A. Sirlin, \prv22 (1980) 2095;
\nup189 (1981) 442;\nextline
C. H. Llewellyn-Smith, G. G. Ross and J. Wheater , \nup177 (1981) 263.}}
\def\BEGN{\rrr\BEGN{A. Buras, J. Ellis, M. K. Gaillard and D. Nanopoulos,
\nup195 (1978) 66.}}
\def\SODI{\rrr\SODI{H. Georgi, unpublished;\nextline
H. Fritzsch and P. Minkowski, Ann.Phys. 93 (1975) 193;\nextline
H. Georgi and D. Nanopoulos, \nup159 (1979) 59.}}
\def\SOAI{\rrr\SOAI{H. Georgi and D. Nanopoulos, \nup159 (1979) 16;
\nextline
F. del Aguila and L. E. Ib\' a\~ nez, \nup177 (1981) 60. }}
\def\SAKH{\rrr\SAKH{A.D. Sakharov, Pisma ZhETF, 5 (1967) 32.}}
\def\KUZ{\rrr\KUZ{V. A. Kuzmin, Pisma ZhETF, 13 (1970) 335.}}
\def\OKUN{\rrr\OKUN{L. Okun and Y. Zeldovich, Comm.Nucl.Part.Phys.
6 (1976) 69.}}
\def\IGNAT{\rrr\IGNAT{A. Y. Ignatiev, N. Krasnikov, V. Kuzmin
and A. Takhelidze, \plb76 (1978) 436;\nextline
M. Yoshimura, Phys.Rev.Lett. 41 (1978) 281;\nextline
S. Dimopoulos and L. Susskind, \prv18 (1978) 4500;\nextline
S. Weinberg, Phys.Rev.Lett. 42 (1979) 850.}}
\def\ARNO{\rrr\ARNO{P. Arnold and L. McLerran, \prv37 (1988) 1020;
\nextline
S. Khlebnikov and M. Shaposnikov, \nup308 (1988) 885;\nextline
D. Grigoriev, V. Rubakov and M. Shaposnikov, \plb216 (1989) 172.}}
\def\MAJOR{\rrr\MAJOR{G. Gelmini and M. Roncadelli, \plb99 (1981) 411;
\nextline
H. Georgi et al., \nup193 (1983) 297.}}
\def\VALL{\rrr\VALL{ For a review see e.g. J. Valle, Valencia preprint
FTUV/90-36, to appear in Progress in Particle and Nuclear Physics.}}
\def\RAMO{\rrr\RAMO{M. Gell-Mann, P. Ramond and R. Slansky, in
Supergravity, eds. P. van Nieuwenhuizen and D. Freedman
(North-Holland 1979) p. 315; \nextline
T. Yanagida, Proceedings of the Workshop on Unified Theories and the
Baryon Number of the Universe, KEK, Japan (1979).}}
\def\PONT{\rrr\PONT{B. Pontecorvo, Sov.Phys. JETP 6 (1958) 429;
ibid. 7 (1958) 172.}}
\def\OSCIL{\rrr\OSCIL{L. Moscoso, Saclay preprint DPhPE 90-14 (1990),
to appear in the Proceedings of "Neutrino 90", CERN, June 1990.}}
\def\DAVI{\rrr\DAVI{R. Davis, D. Harmer and K. Hoffman, Phys.Rev.Lett.
20 (1968) 1205;\nextline
J. Rowley, B. Cleveland and R. Davis, in "Solar Neutrinos and
Neutrino Astronomy, edited by M. L. Cherry;\nextline
K. Lande, "Neutrino 90", Proceedings of the 14th Int. Conf. on
Neutrino Physics and Astrophysics. Editor K. Winter
(North-Holland), in press.}}
\def\KAMIO{\rrr\KAMIO{K. S. Hirata et al. Phys.Rev.Lett. 63 (1989) 16;
Phys.Rev.Lett. 65 (1990) 1301; KEK preprint 90-43 (1990);\nextline
A. Suzuki, KEK preprint 9-31 (1990).}}
\def\SSM{\rrr\SSM{J. N. Bahcall and R. K. Ulrich, Rev.Mod.Phys.
60 (1988) 297.}}
\def\MSW{\rrr\MSW{S. Mikheyev and A. Smirnov, Sov.J.Nucl.Phys. 42
(1985) 913;\nextline
L. Wolfenstein, \prv17 (1978) 2369; \prv20 (1979) 2634;\nextline
For a review see:\nextline
S. Mikheyev and A. Smirnov, Sov.Phys.Usp. 30 (1987) 759.}}
\def\GALL{\rrr\GALL{V.N. Gavrin in "Neutrino 90", CERN, June 1990;
\nextline
T. Kirsten, ibid.}}
\def\CISN{\rrr\CISN{A. Cisneros, Astrophys.Space Sci. 10 (1981) 87;
\nextline
L. Okun, M. Voloshin and M. Vysotsky, Sov.J.Nucl.Phys. 91 (1986) 754;
Sov. Phys. JETP 64 (1986) 446.}}
\def\BABU{\rrr\BABU{K. Babu and V. Mathur, \plb196 (1987) 218;\nextline
M. Fukugita and T. Yanagida, Phys.Rev.Lett. (1987) 1807.}}
\def\SUSY{\rrr\SUSY{Y. Golfand and E. Likhtman, JETP Lett. 13 (1971) 323;
\nextline
D. Volkov and V. Akulov, Pis'ma Zh.ETF 16 (1972) 621;\nextline
P. Ramond, \prv3 (1971) 2415.}}
\def\WZ{\rrr\WZ{J. Wess and B. Zumino, \nup70 (1974) 139.}}
\def\NILL{\rrr\NILL{For phenomenological discussions of supersymmetry
see:\nextline
H. P. Nilles, Phys.Rep. C110 (1984) 1;\nextline
H. Haber and G. Kane, Phys.Rep. C117 (1985) 75;\nextline
G.G. Ross, `Grand Unified Theories' (Benjamin, New York, 1984);\nextline
S.Ferrara ed., `Supersymmetry' (2 Vols.), North-Holland-World
Scientific, Singapore (1987).}}
\def\SUHY{\rrr\SUHY{M. Veltman, Acta Phys.Polon. B12 (1981) 437;
\nextline
L. Maiani, Proceedings of the Summer School of Gif-Sur-Yvette
(Paris 1980).}}
\def\FAY{\rrr\FAY{P. Fayet, \plb69 (1977) 489;\plb84 (1979) 416;
\plb78 (1978) 417;\nextline
G. Farrar and P. Fayet, \plb79 (1978) 442;\plb89 (1980) 191.}}
\def\DRW{\rrr\DRW{S. Dimopoulos, S. Raby and F. Wilczek, \prv24 (1981)
1681.}}
\def\IRA{\rrr\IRA{L. E. Ib\' a\~nez and G. G. Ross, \plb105 (1982) 439;
\nextline
M. Einhorn and D. R. T. Jones, \nup196 (1982) 475.}}
\def\LANGS{\rrr\LANGS{P. Langacker, Pennsylvania preprint UPR-0435T
(1990).}}
\def\WEIN{\rrr\WEIN{S. Weinberg, \prv26 (1982) 287;\nextline
N. Sakai and T. Yanagida, \nup197 (1982) 533.}}
\def\SPD{\rrr\SPD{S. Dimopoulos, S. Raby and F. Wilczek, \plb112
(1982) 133;\nextline
J. Ellis, D. Nanopoulos and S. Rudaz, \nup202 (1982) 43.}}
\def\CFGVP{\rrr\CFGVP{E. Cremmer, S. Ferrara, L. Girardello and A.
Van Proeyen, \nup212 (1983) 413.}}
\def\IBA{\rrr\IBA{L. E. Ib\'a\~nez, \plb118 (1982) 73; \nup218
(1983) 514.}}
\def\BFS{\rrr\BFS{R. Barbieri, S. Ferrara and C. Savoy, \plb119
(1982) 343;\nextline
P. Nath, R. Arnowitt and A. Chamseddine, \plb49 (1982) 970.}}
\def\IBB{\rrr\IBB{L. E. Ib\'a\~nez, \nup218 (1983) 514.}}
\def\IBLO{\rrr\IBLO{L. E. Ib\' a\~ nez and C. L\' opez, \plb126
(1983) 54; \nup233 (1984) 511.}}
\def\AGPW{\rrr\AGPW{L. Alvarez-Gaum\' e, J. Polchinsky and M. Wise,
\nup221 (1983) 495.}}
\def\EHNT{\rrr\EHNT{J. Ellis, J. Hagelin, D. Nanopoulos and K. Tamvakis,
\plb125 (1983) 275.}}
\def\HPNI{\rrr\HPNI{H. P. Nilles, \plb115 (1982) 193.}}
\def\NSW{\rrr\NSW{H. P. Nilles, M. Srednicki and D. Wyler,
\plb120 (1983) 275.}}
\def\GIM{\rrr\GIM{J. Ellis and D. Nanopoulos, \plb110 (1982) 211;
\nextline
R. Barbieri and R. Gatto, \plb110 (1982) 211;\nextline
T. Inami and C. S. Lim, \nup207 (1982) 593.}}
\def\IRB{\rrr\IRB{L. E. Ib\' a\~ nez and G. G. Ross, \plb110 (1982)
215.}}
\def\SNEU{\rrr\SNEU{L. E. Ib\'a\~nez, \plb137 (1984) 160;\nextline
J. Hagelin, G. Kane and S. Raby, \nup241 (1984) 638.}}
\def\INOU{\rrr\INOU{K. Inoue et al., Prog.Theor.Phys. 67 (1982) 1859.}}
\def\INOUB{\rrr\INOUB{K. Inoue et al., Prog.Theor.Phys. 68 (1982) 927.}}
\def\ILM{\rrr\ILM{L. E. Ib\'a\~nez, C. L\'opez and C. Mu\~ noz,
\nup256 (1985) 218.}}
\def\JRR{\rrr\JRR{S. Jones and G. G. Ross, \plb155 (1984) 69;\nextline
C. Kounnas, A. Lahanas, D. Nanopoulos and M. Quiros,
\nup236 (1984) 438;\nextline
A. Bouquet, J. Kaplan and C. Savoy, \nup262 (1985) 299.}}
\def\FLSH{\rrr\FLSH{R. Flores and M. Sher, Ann.Phys. 148 (1983) 95;
\nextline
H. P. Nilles and M. Nussbaumer, \plb145 (1984) 73;\nextline
P. Majumdar and P. Roy, \prv30 (1984) 2432.}}
\def\HAHE{\rrr\HAHE{H. E. Haber and R. Hempfling, Phys.Rev.Lett. 66
(1991) 1815       ;\nextline
J. Ellis, G. Ridolfi and F. Zwirner, \plb257 (1991) 83; \plb262 (1991)
477;\nextline
Y. Okada, M. Yamaguchi and T. Yanagida, Prog.Theor.Phys. Lett. 85 (1991)
1;\plb262 (1991) 54;\nextline
R. Barbieri and M. Frigeni, \plb258 (1991) 395;\nextline
R. Barbieri, F. Caravaglios and M. Frigeni, \plb258 (1991) 167;\nextline
J.R. Espinosa and M. Quiros, \plb266 (1991) 389.}}

\def\EIR{\rrr\EIR{J. Ellis and G. G. Ross, \plb117 (1982) 397;
\nextline
J. Ellis, L. E. Ib\' a\~ nez and G. G. Ross, \nup221 (1983) 445;
\nextline
A. Chamseddine, P. Nath and R. Arnowitt, \plb129 (1983) 445;
\nextline
P. Dicus, S. Nandi and X. Tata, \plb129 (1983) 451.}}
\def\CDFMT{\rrr\CDFMT{CDF Collaboration (presented by G. P. Yeh),
Fermilab-Conf-90/138-E \nextline (1990).}}
\def\CASCA{\rrr\CASCA{H. Baer, X. Tata and J. Woodside,
Phys.Rev.Lett. 63 (1989) 352;\prv41 (1990) 906;\nextline
H. Baer, D. Karatas and X. Tata, Florida State University
preprint FSU-HEP-900430.}}
\def\ALBA{\rrr\ALBA{C. Albajar, C. Fuglesang, S. Hellman, F. Pauss
and G. Polesello, Proceedings of the LHC-Workshop, Aachen (1990).
CERN 90-10 (1990).}}
\def\SFER{\rrr\SFER{L3 Collaboration, B. Adeva et al.,\plb233 (1989) 530;
\nextline
ALEPH Collaboration, D. Decamp et al., \plb236 (1990) 86;\nextline
OPAL Collaboration, M. Z. Akrawy et al., \plb240 (1990) 261;\nextline
DELPHI Collaboration, P. Abreu et al., \plb247 (1990) 157.}}
\def\DELPHI{\rrr\DELPHI{DELPHI Collaboration, P. Abreu et al.,
CERN-EP/90-80 (1990).}}
\def\NEUT{\rrr\NEUT{ALEPH Collaboration, D. Decamp et al.,
CERN-EP/90-63 (1990);\nextline
DELPHI Collaboration, P. Abreu et al., CERN-EP/90-80;\nextline
OPAL Collaboration, M. Z. Akrawy et al., CERN-PPE/90-95;\nextline
MARK-II Collaboration, S. Komamiya et al., Phys.Rev.Lett. 64 (1990)
2984.}}
\def\BGGR{\rrr\BGGR{R. Barbieri, G. Gamberini, G. Giudice and
G. Ridolfi, \plb195 (1987) 500; \nup296 (1988) 75;\nextline
A. Bartl et al., preprints HEPHY-PUB 526/89 and UWThPh-1989-38\nextline
(1989);\nextline
J. Ellis, G. Ridolfi and F. Zwirner, \plb237 (1990) 423.}}
\def\SHIG{\rrr\SHIG{ALEPH Collaboration, D. Decamp et al., \plb237
(1990) 291;\nextline
DELPHI Collaboration, P. Abreu et al., \plb245 (1990) 276;\nextline
OPAL Collaboration, M. Z. Akrawy et al., CERN-EP/90-100 (1990);
\nextline
L3 Collaboration, B. Adeva et al., CERN-PPE-L3-015 (1990);\nextline
MARK-II Collaboration, S. Komamiya et al. Phys.Rev.Lett. 64 (1990)
2881.}}
\def\HHG{\rrr\HHG{For a useful "Higgs-hunting guide" see:
S. Dawson, J. Gunion, H. Haber and G. Kane, BNL-41644 (1989)
(to appear in Phys.Rep.).}}
\def\IM{\rrr\IM{L. E. Ib\' a\~ nez and J. Mas, \nup286 (1987) 107.}}
\def\DS{\rrr\DS{J. P. Derendinger and C. Savoy, \nup237 (1984) 307.}}
\def\DREE{\rrr\DREE{M. Drees, Int.Jour.Mod.Phys.A 4 (1989) 3635;
\nextline
J. Ellis, J. Gunion, H. Haber, L. Roszkowski and F. Zwirner,
Phys.Rev. D39 (1989) 844;\nextline
P. Binetruy and C. Savoy, Saclay preprint SPhT/91-143.}}

\def\HALL{\rrr\HALL{L. Hall and M. Suzuki, \nup231 (1984) 419;
\nextline
I. Lee, \nup246 (1984)) 120.}}
\def\ZWI{\rrr\ZWI{F. Zwirner, \plb132 (1983) 103;\nextline
R. Barbieri and A. Masiero, \nup267 (1986) 679.}}
\def\DIMH{\rrr\DIMH{S. Dimopoulos and L. Hall, \plb196 (1987) 135;
\plb207 (1988) 216;\nextline
S. Dimopoulos et al. \prv41 (1990) 2099.}}
\def\HARA{\rrr\HARA{L. Hall and L. Randall, LBL-28879 (1990).}}
\def\BG{\rrr\BG{R. Barbieri and G. Giudice, \nup296 (1988) 75.}}
\def\GSW{\rrr\GSW{M. Green, J. Schwarz and E. Witten, "Superstring
Theory", Vols I and II, Cambridge University Press (1986) ;\nextline
D. Gross, in Proceedings of the 1986 ASI School (Virgin Islands),
Plenum Press (1987).}}
\def\GS{\rrr\GS{M. Green and J. Schwarz, \nup255 (1985) 93.}}
\def\HETE{\rrr\HETE{D. Gross, J. Harvey, E. Martinec and R. Rohm,
\nup256 (1985); \nup267 (1986) 75.}}
\def\ORBI{\rrr\ORBI{L. Dixon, J. Harvey, C. Vafa and E. Witten,
\nup261 (1985) 651;\nextline
L. E. Ib\' a\~nez, H. P. Nilles and F. Quevedo, \plb187 (1987) 25;
\nextline
K. Narain, M. Sarmadi and C. Vafa, \nup288 (1987) 951;\nextline
L. E. Ib\' a\~ nez, J. Mas, H. P. Nilles and F. Quevedo,
\nup301 (1988) 157.}}
\def\KLT{\rrr\KLT{H. Kawai, D. Lewellen and S. Tye, Phys.Rev.Lett.
57 (1986) 1832; \nup288 (1987 )1;\nextline
I. Antoniadis, C. Bachas and C. Kounnas, \nup289 (1987) 87.}}
\def\LLS{\rrr\LLS{W. Lerche, D. L\" ust and A. N. Schellekens,
\nup287 (1987) 477.}}
\def\GEP{\rrr\GEP{D. Gepner, \nup296 (1987) 757;\nextline
Y. Kazama and H. Suzuki, \nup321 (1989) 232.}}
\def\TWIS{\rrr\TWIS{A. Font, L. E. Ib\' a\~nez, F. Quevedo and
A. Sierra, \nup337 (1990) 119.}}
\def\GKMR{\rrr\GKMR{B. Greene, K. Kirklin, P. Miron and G. G. Ross,
\nup278 (1986) 667; \nup279 (1986) 574.}}
\def\FENO{\rrr\FENO{L. E. Ib\' a\~ nez, J. E. Kim, H. P. Nilles
and F. Quevedo, \plb191 (1987) 282;\nextline
A. Font, L. E. Ib\' a\~ nez, F. Quevedo and A. Sierra, \nup331 (1990)
421.}}
\def\FLIP{\rrr\FLIP{I. Antoniadis, J. Ellis, J. Hagelin and D. Nanopoulos
, \plb231 (1989) 65 and references therein.}}
\def\GINS{\rrr\GINS{P. Ginsparg, \plb197 (1987) 139.}}
\def\SHE{\rrr\SHE{A. N. Schellekens, \plb237 (1990) 363.}}
\def\HOSO{\rrr\HOSO{Y. Hosotani, \plb129 (1983) 193;\nextline
E. Witten, \nup327 (1989) 673.}}
\def\LLEW{\rrr\LLEW{D. Lewellen, \nup337 (1990) 61;\nextline
J.A. Schwarz, Phys.Rev.D42 (1990) 389.}}
\def\LEVEL{\rrr\LEVEL{ A. Font, L. E. Ib\' a\~ nez and F. Quevedo,
\nup345 (1990) 389.}}
\def\MISS{\rrr\MISS{S. Dimopoulos and F. Wilczek, Santa Barbara
preprint (1981); Proc. Erice Summer School (1981);\nextline
B. Grinstein, \nup206 (1982) 387;\nextline
A. Masiero, D. Nanopoulos, K. Tamvakis and T. Yanagida, \plb115
(1982) 380.}}
\def\BD{\rrr\BD{T. Banks and L. Dixon, \nup307 (1988) 93.}}
\def\DIN{\rrr\DIN{J. P. Derendinger, L. E. Ib\' a\~ nez and H. P. Nilles,
\plb155 (1985) 65;\nextline
M. Dine, R. Rohm, N. Seiberg and E. Witten, \plb156 (1985) 55.}}
\def\ANOM{\rrr\ANOM{M. Green and J. Schwarz, \plb149 (1984) 117;\nextline
W. Lerche, B. Nilsson and A. N. Schellekens, \nup299 (1988) 91;\nextline
J. A. Casas, E. Katehou and C. Mu\~ noz, \nup317 (1989) 171.}}
\def\FAYIL{\rrr\FAYIL{M. Dine, N. Seiberg and E. Witten, \nup289 (1987)
585;\nextline
J. Atick, L. Dixon and A. Sen, \nup292 (1987) 109;\nextline
M. Dine, I. Ichinose and N. Seiberg, \nup293 (1987) 253.}}
\def\REAL{\rrr\REAL{A. Font, L. E. Ib\' a\~ nez, H. P. Nilles and
F. Quevedo, \plb210 (1988) 101; \plb213 (1988) 564;\nextline
J. A. Casas and C. Mu\~ noz, \plb209 (1988) 214; \plb214 (1988) 63.}}
\def\FGN{\rrr\FGN{H. P. Nilles, \plb115 (1982) 193;\nextline
S. Ferrara, L. Girardello and H. P. Nilles, \plb125 (1983) 457.}}
\def\IN{\rrr\IN{L. E. Ib\' a\~ nez and H. P. Nilles, \plb169 (1986) 354;
\nextline
T. Taylor and G. Veneziano, \plb212 (1988) 147;
\nextline
L. Dixon, V. Kaplunovsky and J. Louis, SLAC-PUB-5138 (1990).}}
\def\SEMI{\rrr\SEMI{A. Font, L. E. Ib\' a\~ nez, D. L\" ust and
F. Quevedo, \plb245 (1990) 401;\nextline
S. Ferrara, N. Magnoli, T. Taylor and G. Veneziano, \plb245 (1990) 409.}}
\def\OLEC{\rrr\OLEC{H. P. Nilles and M. Olechowski, \plb248 (1990)
268;\nextline
P. Binetruy and M. K. Gaillard, preprint  CERN-TH.
5727/90.}}
\def\DGSB{\rrr\DGSB{L.E. Ib\'a\~nez and G.G. Ross, \nup368 (1992) 3.}}
\def\ILB{\rrr\ILB{L.E. Ib\'a\~nez and D. L\"ust, CERN-TH.6380 (1992).}}
\def\REFIN{\rrr\REFIN{U. Ellwanger, \nup238 (1984) 665 ;\nextline
H.J.Kappen, Phys.Rev.D38 (1988) 721;\nextline
G. Gamberini, G. Ridolfi and F. Zwirner, \nup331 (1990) 331.}}
\def\GIRAR{\rrr\GIRAR{L. Girardello and M. Grisaru, \nup194 (1982) 65.}}
\def\DREEB{\rrr\DREEB{M. Drees and M.M. Nojiri, KEK preprint
KEK-TH-290 (1991).}}
\def\BAHALL{\rrr\BAHALL{R. Barbieri and L. Hall, LBL preprint
LBL-31238 (1991).}}
\def\RR{\rrr\RR{R.G. Roberts and G.G. Ross, Rutherford Lab. preprint
RAL-92-005 (1991).}}
\def\DA{\rrr\DA{M. Daniel and J.A. Pe\~narrocha, \plb127 (1983) 219.}}

\def\PLANCK{\rrr\PLANCK{L.E. Ib\'a\~nez, \plb126 (1983) 196;
\nextline J.E. Bjorkman and D.R.T. Jones, \nup259 (1985) 533.}}

\def\ZNZM{\rrr\ZNZM{A. Font, L.E. Ib\'a\~nez and F. Quevedo,
\plb217 (1989) 272.}}

\def\IN{\rrr\IN{L.E. Ib\'a\~nez and H.P. Nilles, \plb169 (1986) 354.}}

\def\EINHJON{\rrr\EINHJON{M. Einhorn and D.R.T. Jones, \nup196 (1982)
                      475.}}

\def\KAPLU{\rrr\KAPLU{V. Kaplunovsky, \nup307 (1988) 145.}}

\def\AMALDI{\rrr\AMALDI{
J. Ellis, S. Kelley and D.V. Nanopoulos, \plb249 (1990) 441;
\plb260 (1991) 131; \nextline
P. Langacker,      ``Precision tests of the standard model''
Pennsylvania preprint UPR-0435T, (1990);\nextline
U. Amaldi, W. de Boer and H. F\"urstenau, \plt260 (1991) 447;\nextline
P. Langacker and M. Luo, Phys.Rev.D44 (1991) 817;\nextline
R.G. Roberts and G.G. Ross, talk presented by G.G. Ross at 1991 Joint
International Lepton-Photon Symposium and EPS Conference,     to
be published.
}}

\def\DG{\rrr\DG{S. Dimopoulos, S. Raby and F. Wilczek, Phys. Rev.
D24 (1981) 1681;\nextline
                 L.E. Ib\'a\~nez and G.G. Ross, \plb105 (1981) 439;
\nextline S. Dimopoulos and H. Georgi, \nup193 (1981) 375.}}

\def\IMNQ{\rrr\IMNQ{L.E. Ib\'a\~nez, H.P. Nilles and F. Quevedo,
\plt187 (1987) 25; L.E. Ib\'a\~nez, J. Mas, H.P. Nilles and
F. Quevedo, \nup301 (1988) 157; A. Font, L.E. Ib\'a\~nez,
F. Quevedo and A. Sierra, \nup331 (1990) 421.}}

\def\SCHELL{\rrr\SCHELL{A.N. Schellekens, \plt237 (1990) 363.}}

\def\GQW{\rrr\GQW{H. Georgi, H.R. Quinn and S. Weinberg, Phys. Rev.
Lett. ${\underline{33}}$ (1974) 451.}}

\def\GINS{\rrr\GINS{P. Ginsparg, \plt197 (1987) 139.}}

\def\ELLISETAL{\rrr\ELLISETAL{I. Antoniadis, J. Ellis, R. Lacaze
and D.V. Nanopoulos, {\it ``String Threshold Corrections and
                Flipped $SU(5)$'',} preprint CERN-TH.6136/91 (1991);
    S. Kalara, J.L. Lopez and D.V. Nanopoulos, {\it``Threshold
   Corrections and Modular Invariance in Free Fermionic Strings'',}
      preprint CERN-TH-6168/91 (1991).}}

\def\DFKZ{\rrr\DFKZ{J.P. Derendinger, S. Ferrara, C. Kounnas and
F. Zwirner,{\it ``On loop corrections to string effective field theories:
         field-dependent gauge couplings and sigma-model anomalies'',}
        preprint CERN-TH.6004/91, LPTENS 91-4 (revised version) (1991).}}

\def\LOUIS{\rrr\LOUIS{J. Louis, {\it
         ``Non-harmonic gauge coupling constants in supersymmetry
         and superstring theory'',} preprint SLAC-PUB-5527 (1991);
         V. Kaplunovsky and J. Louis, as quoted in J. Louis,
         SLAC-PUB-5527 (1991).}}

\def\DIN{\rrr\DIN{J.P. Derendinger, L.E. Ib\'a\~nez and H.P Nilles,
        \nup267 (1986) 365.}}

\def\DHVW{\rrr\DHVW{L. Dixon, J. Harvey, C.~Vafa and E.~Witten,
         \nup261 (1985) 651;
        \nup274 (1986) 285.}}

\def\DKLB{\rrr\DKLB{L. Dixon, V. Kaplunovsky and J. Louis,
         \nup355 (1991) 649.}}

\def\DKLA{\rrr\DKLA{L. Dixon, V. Kaplunovsky and J. Louis, \nup329 (1990)
            27.}}

\def\ALOS{\rrr\ALOS{E. Alvarez and M.A.R. Osorio, \prv40 (1989) 1150.}}

\def\FILQ{\rrr\FILQ{A. Font, L.E. Ib\'a\~nez, D. L\"ust and F. Quevedo,
           \plt245 (1990) 401.}}

\def\CFILQ{\rrr\CFILQ{M. Cvetic, A. Font, L.E.
           Ib\'a\~nez, D. L\"ust and F. Quevedo, \nup361 (1991) 194.}}

\def\FILQ{\rrr\FILQ{A. Font, L.E. Ib\'a\~nez, D. L\"ust and F. Quevedo,
           \plt245 (1990) 401.}}

\def\DUAGAU{\rrr\DUAGAU{S. Ferrara, N. Magnoli, T.R. Taylor and
           G. Veneziano, \plt245 (1990) 409;\nextline  H.P. Nilles and M.
           Olechowski, \plt248 (1990) 268;\nextline  P. Binetruy and M.K.
           Gaillard, \plt253 (1991) 119;\nextline J. Louis, SLAC-PUB-5645
           (1991);\nextline S. Kalara, J. Lopez and D. Nanopoulos,
           Texas A\&M  preprint CTP-TAMU-69/91.}}

\def\CFILQ{\rrr\CFILQ{M. Cvetic, A. Font, L.E.
           Ib\'a\~nez, D. L\"ust and F. Quevedo, \nup361 (1991) 194.}}

\def\FIQ{\rrr\FIQ{A. Font, L.E. Ib\'a\~nez and F. Quevedo,
        \plt217 (1989) 272.}}

\def\FLST{\rrr\FLST{S. Ferrara,
         D. L\"ust, A. Shapere and S. Theisen, \plt225 (1989) 363.}}

\def\FLT{\rrr\FLT
{S. Ferrara, D. L\"ust and S. Theisen, \plt233 (1989) 147.}}

\def\IBLU{\rrr\IBLU{L. Ib\'a\~nez and D. L\"ust,
          \plb267 (1991) 51.}}

\def\GAUGINO{\rrr\GAUGINO{J.P. Derendinger, L.E. Ib\'a\~nez and H.P. Nilles,
            \plb155 (1985) 65;
         M. Dine, R. Rohm, N. Seiberg and E. Witten, \plb156 (1985) 55.}}

\def\GHMR{\rrr\GHMR{D.J. Gross, J.A. Harvey, E. Martinec and R. Rohm,
         \prl54 (1985) 502; \nup256 (1985) 253; \nup267 (1986) 75.}}

\def\ANT{\rrr\ANT{I. Antoniadis, K.S. Narain and T.R. Taylor,
        \plb267 (1991) 37.}}

\def\WITTEF{\rrr\WITTEF{E. Witten, \plb155 (1985) 151.}}

\def\HB{\rrr\HB{L. Hall and R. Barbieri, private communication and
preprint in preparation (1991).}}

\def\SCHELL{\rrr\SCHELL{A.N. Schellekens, ``Superstring
construction'', North Holland, Amsterdam (1989).}}

\def\FLIP{\rrr\FLIP{I. Antoniadis, J. Ellis, J. Hagelin and
D.V. Nanopoulos, \plb231 (1989) 65 and references therein. For a
recent review see J. Lopez and D.V. Nanopoulos,
                  Texas A\& M preprint CTP-TAMU-76/91 (1991).}}

\def\GKMR{\rrr\GKPM{B. Greene, K. Kirklin, P. Miron and G.G. Ross,
\nup292 (1987) 606.}}

\def\OTH{\rrr\OTH{D. Bailin, A. Love and S. Thomas, \plb188 (1987)
193; \plb194 (1987) 385; B. Nilsson, P. Roberts and P. Salomonson,
\plb222 (1989) 35;
J.A. Casas, E.K. Katehou and C. Mu\~noz, \nup317 (1989) 171;
J.A. Casas and C. Mu\~noz, \plb209 (1988) 214, \plb212 (1988) 343
J.A. Casas, F. Gomez and C. Mu\~noz, \plb251 (1990) 99;
A. Chamseddine and J.P. Derendinger, \nup301 (1988) 381;
A. Chamseddine and M. Quiros, \plb212 (1988) 343, \nup316 (1989) 101;
T. Burwick, R. Kaiser and H. M\"uller, \nup355 (1991) 689;
Y. Katsuki, Y. Kawamura, T. Kobayashi, N. Ohtsubo,
Y. Ono and K. Tanioka, \nup341 (1990) 611.}}

\def\SCHLUS{\rrr\SCHLUS{For a review, see e.g. J. Schwarz, Caltech
preprint CALT-68-1740 (1991); D. L\"ust,  CERN preprint TH.6143/91.
}}

\def\LMN{\rrr\LMN{J. Lauer, J. Mas and H.P. Nilles, \plb226 (1989)
251, \nup351 (1991) 353; W. Lerche, D. L\"ust and N.P. Warner,
\plb231 (1989) 417.}}

\def\ILR{\rrr\ILR{ L.E. Ib\'a\~nez, D. L\"ust and G.G. Ross,
\plb272 (1991) 251.}}

\def\ANTON{\rrr\ANTON{I. Antoniadis, J. Ellis, S. Kelley and
D.V. Nanopoulos, \plb272 (1991) 31 .}}

\def\HIGH{\rrr\HIGH{D. Lewellen, \nup337 (1990) 61;
J.A. Schwartz, Phys.Rev. D42 (1990) 1777.}}

\def\HIGHK{\rrr\HIGHK{A. Font, L.E. Ib\'a\~nez and F. Quevedo,
\nup345 (1990) 389; J. Ellis, J. Lopez and D.V. Nanopoulos,
\plb245 (1990) 375.}}

\def\LLR{\rrr\LLR{C.H. Llewellyn-Smith, G.G. Ross and
J.F. Wheater, \nup177 (1981) 263; \nextline
S. Weinberg, \plb91 (1980) 51; \nextline
L. Hall, \nup178 (1981) 75;\nextline
P. Binetruy and T. Schucker, \nup178 (1981) 293.}}

\def\IL{\rrr\IL{  L.E. Ib\'a\~nez and D. L\"ust, CERN-TH.6380/92
(1992).}}

\def\YO{\rrr\YO{ L.E. Ib\'a\~nez, {\it``Some topics
in the low energy physics from superstrings''} in proceedings of
the NATO workshop on ``Superfield Theories", Vancouver,
Canada. Plenum Press, New York (1987).}}

\def\SDUAL{\rrr\SDUAL{A. Font, L.E. Ib\'a\~nez, D. L\"ust
and F. Quevedo, \plb249 (1990) 35.}}

\def\MALLOR{\rrr\MALLOR{For a recent review see L.E. Ib\'a\~nez,
{\it ``Beyond the Standard Model (yet again)''}, CERN preprint
TH.5982/91, to appear in the Proceedings of the 1990 CERN
School of Physics, Mallorca (1990).}}

\def\FIQS{\rrr\FIQS{A. Font, L.E. Ib\'a\~nez, F. Quevedo and
A. Sierra, \nup337 (1990) 119.}}

\def\LT{\rrr\LT{D. L\"ust and T.R. Taylor, \plb253 (1991) 335;
B. Carlos, J. Casas and C. Mu\~noz, preprint CERN-TH.6049/91
(1991).}}

\def\DGSA{\rrr\DGSA{L.E. Ib\'a\~nez and G.G. Ross, \plb260 (1991) 291.}}

\def\HLW{\rrr\HLW{L. Hall, J. Lykken and S. Weinberg, Phys.Rev.D27 (1983)
  2359.}}

\def\MBMT{\rrr\MBMT{ G. Lazarides and Q. Shafi, Bartol Research
preprint BA-91-25 (1991);\nextline
S. Kelley, J. L\'opez and D. Nanopoulos, Texas AM preprint
CTP-TAMU-79-91 (1991);\nextline
H. Aranson, D. Casta\~no, B. Keszthelyi, S. Mikaelian, E. Piard,
P. Ramond and B. Wright, Phys.Rev.Lett. 67 (1991) 2933;\nextline
S. Dimopoulos, L. Hall and S. Raby, LBL-31441 UCB-PTH 91/61.}}

\def\SUSYGUT{\rrr\SUSYGUT{E. Witten, \nup188 (1981) 513;\nextline
S. Dimopoulos and H. Georgi, \nup193 (1981) 150;\nextline
N. Sakai, Z.Phys. C11 (1982) 153;\nextline
E. Witten, \plb105 (1981) 267;\nextline
L.E. Ib\'a\~nez and G.G. Ross, \plb105 (1981) 439; \plb110 (1982) 215
\nextline
L. Alvarez-Gaum\'e, M. Claudson and M. Wise, \nup207 (1982) 16;\nextline
M. Dine and W. Fischler, \nup204 (1982) 346;\nextline
J. Ellis, L.E. Ib\'a\~nez and G.G. Ross, \plb113 (1982) 283;\nup221
(1983)  29;\nextline
C. Nappi and B. Ovrut, \plb113 (1982) 175;\nextline
S. Dimopoulos and S. Raby, \nup219 (1983) 479;\nextline
J. Polchinski and L. Susskind, Phys.Rev.D26 (1982) 3661;\nextline
J. Ellis, D. Nanopoulos and K. Tamvakis \plb121 (1983) 123;\nextline
H.P. Nilles, \nup217 (1983) 366.}}

\def\AAA{\rrr\AAA{J.M. Frere, D.R.T. Jones and S. Raby, \nup222 (1983)
11.}}

\def\AAB{\rrr\AAB{M. Drees, M. Gl\"uck and K. Grassie, \plb157 (1985)
164.}}

\def\OFFSET{\hoffset=12.pt\voffset=55.pt}
\def\SIZE{\hsize=420.pt\vsize=620.pt}

\catcode`@=12
\newtoks\Pubnumtwo
\newtoks\Pubnumthree
\catcode`@=11
\def\p@bblock{\begingroup\tabskip=\hsize minus\hsize
   \baselineskip=0.5\ht\strutbox\topspace-2\baselineskip
   \halign to \hsize{\strut ##\hfil\tabskip=0pt\crcr
   \the\Pubnum\cr  \the\Pubnumtwo\cr 
   \the\pubtype\cr}\endgroup}
\pubnum={6412/92}
\date{February  1992}
\pubtype={}
\titlepage
\vskip -.6truein
\title{\bf   Electroweak Breaking in Supersymmetric Models* }
 \centerline{\bf Luis E. Ib\'a\~nez**}
 \vskip .1truein
  \centerline{CERN, 1211 Geneva 23, Switzerland}
\vskip 0.1truein
\centerline{and}
\vskip 0.1truein
\centerline{\bf Graham G. Ross}
\vskip .1truein
\centerline{Dep. Theoretical Physics, Oxford University, England}
\abstract\noindent\nobreak
We discuss the mechanism for electroweak symmetry breaking in
supersymmetric versions of the standard model. After briefly
reviewing the possible sources of supersymmetry breaking, we
show how the required pattern of symmetry breaking can
automatically result from the structure of quantum corrections
in the theory. We demonstrate that this radiative breaking
mechanism works well for a heavy top quark and can be combined
in unified versions of the theory with excellent predictions for
the running couplings of the model.

\vskip 1.0cm

\bigskip
    {\bf *} \ To appear in ``Perspectives in Higgs Physics'',
G. Kane editor.
\bigskip
         {\bf **} \ Adress after $1^{st}$ October 1992: Departamento
de F\'isica Te\'orica C-XI, Universidad Aut\'onoma de Madrid,
Cantoblanco, 28049 Madrid, Spain.
\endpage
\pagenumber=1
\sequentialequations

\leftline{\bf  1\ Introduction}

The principle motivation which suggests the standard model should
be extended to make it supersymmetric through the addition of new
light supersymmetric states, accessible to particle accelerators,
is the need to solve the hierarchy problem \GILD  ,\TOOFT ,\SUHY   .
This is the problem
of explaining why the W and Z-boson masses are so light in
comparison with the unification or Planck scale despite the fact
that radiative corrections associated with elementary Higgs
scalar fields typically drive the electroweak breaking scale
close to the largest mass scale in the theory. In this article
we will describe how a low-energy supersymmetric extension of the
standard model \NILL ,\MALLOR\
               solves the hierarchy problem by limiting the
magnitude of the radiative corrections. We will also discuss how
the breaking of the electroweak symmetry appears as a consequence
of the quantum corrections of the theory and show how this may
   easily  be included in a unification scheme.

In the standard model the hierarchy problem is due to the
existence of elementary Higgs scalar fields. These fields have
pointlike couplings associated with their gauge, gravitational
and other interactions which lead to large radiative corrections
to their mass and the expectation that they should have a mass
comparable to the largest mass scale in the theory. As a result
the electroweak breaking scale which is related to the Higgs mass
is also expected to be unacceptably large.  Two solutions to the
hierarchy problem have been suggested. It could be that the Higgs
(and other) fields may not be elementary and the pointlike
interaction only applies up to the scale at which the composite
structure appears. This has the effect of cutting off the
radiative corrections at this composite scale and provided this
is small enough the hierarchy problem is avoided. Examples of
composite solutions to the hierarchy problem will be found
elsewhere in this volume. The second possible solution is that
the Higgs scalars are indeed elementary with pointlike
interactions up to the Planck scale but that a symmetry protects
the Higgs scalars from large radiative corrections to their
masses. It has been shown that the only possible such symmetry
is supersymmetry, hence the interest in low energy supersymmetric
versions of the standard model
\FAY ,\DFS ,                  \SUSYGUT . Although it appeared that there
was considerable freedom in building supersymmetric theories
corresponding to the possibility of several supersymmetry
generators, it was soon clear that the chiral nature of the
standard model probably only allows a supersymmetric extension with
only a single (N=1) supersymmetry.\footnote*{Matter must belong
to vectorlike representations in $N>1$ theories corresponding to
the existence of mirror fermions. The difficulty in making these
(so-far unobserved) states heavy has essentially killed attempts
to build viable models with $N>1$.} Thus we turn now to a
consideration of how to build such an N=1 extension of the
standard model.

\bigskip

\leftline{\bf 1.1 \ The supermultiplet content of the supersymmetric
standard model}

In order to create a supersymmetric version of the standard model
it is necessary to assign its states to N=1 supermultiplets.
These consist of pairs of states differing in helicity,
$\lambda$, by $1/2$        . The gauge bosons of the standard
model are assigned to gauge supermultiplets which contain a
$\lambda=1,1/2        $ helicity pair and thus to every gauge
boson there is assigned a fermion partner called a gaugino. Since
the supersymmetry generator commutes with the Yang Mills
generators the gauge quantum numbers of the gaugino are the same
as its gauge boson partner. The resulting set of gauge
supermultiplets needed is given in Table 1.

\tabskip= 3 pt plus 5 pt
$$\vbox{
\halign to 14.0 true cm{
\hfil $#$ \hfil &
\hfil $#$ \hfil &
\hfil $#$ \hfil &
\hfil $#$ \hfil &
\hfil $#$ \hfil &
\hfil $#$ \hfil\cr
\hbox{VECTOR}&\hbox{MULTIPLETS} &\ &\ &\hbox{CHIRAL} &\hbox{MULTIPLETS}
 \cr
\noalign{\smallskip \hrule \bigskip}
J=1& J=1/2 &  &  & J=1/2 & J=0 \cr
\noalign{\bigskip}
g & {\tilde g} &  &  & Q_L,U_L^c,D_L^c & {\tilde Q}_L,{\tilde U}_L^c,
{\tilde D}_L^c           \cr
\noalign{\bigskip}
W^{\pm },W^0 &{\tilde W}^{\pm },{\tilde W}^0 & &  &L_L,E_L^c&
{\tilde L}_L,{\tilde E}_L^c \cr
\noalign{\bigskip}
B & {\tilde B}& & &{\tilde H}_1,{\tilde H}_2&H_1,H_2  \cr
\noalign{\bigskip}
\noalign{\bigskip \hrule}
}}$$

\centerline{Table 1}

The matter fields of the standard model, the quarks and leptons,
must also be assigned to supermultiplets. Because, in the
standard model, the gauge quantum numbers of the gauge bosons and
gauginos are different from the matter fields it is not possible
to identify any of the latter with any of the gauginos. Thus we
must introduce further supermultiplets and since we cannot allow
additional gauge bosons without enlarging the gauge group we are
forced to assign the matter states to               chiral
supermultiplets which contain a $\lambda=1/2        ,0$ helicity
pair. Thus each quark and each lepton is assigned to a (left-handed)
chiral supermultiplet with a scalar partner, a squark and a slepton
respectively. These are also given in Table 1            and
again the gauge quantum numbers of the scalar states are
identical to those of their fermion partners.

Finally it is necessary to assign the Higgs scalars of the
standard model to a supermultiplet. Although there are colour
singlet, electroweak doublet slepton states that appear to have
the correct quantum numbers to be identified with the Higgs
doublet, this proves to be unsuitable. The reason is that the
constraints of supersymmetry on the Yukawa couplings of the
theory prevent one from coupling the matter fields in the chiral
supermultiplets of Table 1            to the conjugate scalar
fields of the chiral supermultiplets, only coupling to non-
conjugate fields is allowed. As a result it is necessary to have
two (left-handed) chiral superfields containing {\it two} Higgs
doublets, $H_{1,2}$ with opposite hypercharge Y=-1,1 to allow for
the couplings needed to give both up and down quarks a mass. Thus
even if we identify $H_{1}$ with a slepton supermultiplet it is
necessary to introduce an additional chiral supermultiplet to
accommodate $H_{2}$. In this case its fermion partner will
introduce an anomaly coupled to hypercharge. To avoid this it is
necessary to introduce {\it two} additional supermultiplets with
hypercharge Y=1,-1 and then it is usual to identify the two Higgs
doublets with the scalar components of these supermultiplets as
is shown in Table 1           . While this is not unique the
alternative identification of $H_{1}$ with one of the slepton
doublets introduces lepton number violation and is avoided in the
minimal supersymmetric version of the standard model (the MSSM).

\bigskip

\leftline{\bf 1.2  The couplings of the MSSM}
To complete the definition of the MSSM it is necessary to specify
the couplings of the theory. The vector boson gauge couplings to
the new states are all specified since the gauge quantum numbers
of the new supersymmetric partners of the standard model states
have all been specified to be the same as their standard model
partners. Operation by the supersymmetry generator induces new
couplings involving the gaugino partner of the gauge boson with
a strength given in terms of the original gauge coupling. The
resulting Feynman rules are well known \NILL ,\MALLOR\
and we will not reproduce them here.
  In addition to the gauge couplings the MSSM must have the
Yukawa couplings necessary to give mass to the quarks
and leptons. These are associated with new scalar couplings
related by the operation of the supersymmetry generator. The
totality of these terms is most conveniently derived from the
superpotential, P. In order to generate the required Yukawa
couplings P must contain the terms

$$
P =    h_{i j k} L_{i} H_{1j}     {E}_{k} \ +\  h{'}_{ijk}    Q_{i}
H_{1j}     {D}_{k}\ +\  h{''}_{ijk}    Q_{i} H_{2j}     {U}_{k}
\eqn \superp
$$
where $L$ and $E$ ($Q$ and  $U,D$             ) are the (left-handed)
                lepton doublet and antilepton singlet (quark
doublet and antiquark
singlet) chiral superfields respectively and $ H_{1,2}$  are
(left-handed)
Higgs superfields. The supersymmetric couplings correspond to the
F terms of the superpotential $P$. These give both Yukawa couplings
and pure scalar couplings. For example, the Yukawa couplings
following from the first term of
eq.\superp\            are
$$
L_{Yukawa}\ =\ h_{i j k}(L_{i} H_{1j}     {E}_k +{\tilde{L}}_{i}
{\tilde {H}}_{1j}                   E _{k}  +L_{i} {\tilde{H}}_{1j}
         {\tilde{E}}_{k} )
\eqn \yuki
$$
where we denote by a supertwiddle the supersymmetric partners to the
quarks, leptons and Higgs bosons, namely the squarks,    sleptons
and Higgsinos.
The first term is the usual term in the standard model needed to give
charged leptons a mass. The new couplings associated with the
supersymmetric states related to the first term by the operation
of the supersymmetry generator are given by the second and third
terms.
The scalar couplings associated with eq.\superp    \       are

$$
L_{scalar}\ =\   \sum_{i,j} \mid {    {\partial ^{2} P}\over {\partial
\phi_{i} \phi_{j}}} \mid ^2
\eqn \poti
$$
where $\phi_{i}$ are chiral superfields and after differentiation
of the superpotential only the scalar components of the remaining
chiral superfields are kept. The full set of Feynman rules
resulting from the superpotential of eq.\superp           may be
found e.g. in ref.\NILL\  .

There is one further coupling needed to complete the couplings
of the minimal supersymmetric version of the standard model. In
order to generate a mass for the Higgsinos associated with the
Higgs doublets $H_{1,2}$ it is necessary to add a term to the
superpotential given by

$$
P{ '} \     =\ \mu H_{1}H_{2}
\eqn \mumu
$$
In addition to giving a mass $\mu $ to the Higgsinos, this term
plays an important role in determining the Higgs scalar potential
and the pattern of electroweak symmetry breaking.  As we will
discuss in more detail in section 3                 the scalar
term following from eq.\mumu  \  aligns the vacuum expectation
values (vevs) of the two Higgs fields so that the photon is left
massless, obviously a crucial ingredient for a viable theory.

We note that this term is the only one involving a coupling with
dimensions of mass. If the theory is to avoid the hierarchy
problem $\mu $ must be small, of order the electroweak breaking
scale, for the Higgs scalars also get a contribution $\mu ^{2}$
to their mass squared. Thus any complete explanation of the
electroweak breaking scale must explain the origin of $\mu $.
    There is  a simple modification of the theory that avoids the
introduction of a mass at this stage. In this variation of the
MSSM eq.\mumu\   is replaced by

$$
P{'}    \ = \ \lambda H_{1} H_{2} N + \lambda {'}      N^{3}
\eqn \singli
$$
where N is a gauge singlet chiral superfield. As we will see this
term generates a vev, $<N>$, for the scalar component of N which
then gives the superpotential term of eq.\mumu\   with
$\mu=\lambda <N>$. The advantage of starting with eq.\singli\  is
that $<N>$ is automatically driven to be of the same magnitude
as the Higgs vevs, avoiding the need to introduce a new mass
scale by hand.
\endpage

\bigskip
\leftline{\bf  1.3 R-parity and discrete symmetries}
Although we will be primarily concerned with the minimal
supersymmetric standard model defined above it is not the only
way a supersymmetric version of the standard model with minimal
particle content can be constructed. For completeness we add here
a discussion of possible variants of the supersymmetric version
of the standard model. The ambiguity arises because the couplings
of eq.\superp  \      and eq.\mumu \  or eq.\singli\ are not the
only ones allowed by $SU(3)\otimes SU(2)\otimes U(1)$ for we may
add to the superpotential the terms \WEIN\

$$
[\lambda_{i j k} L_{i} L_{j}     {E}_{k} + \lambda'_{i j k} L_{i}
Q_{j}     {D}_{k} + \lambda''_{i j k}     {U}_{i}     {D}_{j}
    {D}_{k}      \ +\ \mu _i{'}L_iH_2 ]_F \
\eqn \rsym
$$

These terms violate baryon  or  lepton  number and, if all are
present in the Lagrangian, they generate via graphs such as in fig. 1
an unacceptably large
amplitude for proton decay suppressed only by the inverse
supersymmetry-breaking
\vskip 3.truein
\centerline{{\bf Fig. 1} Graph giving proton decay in R-parity
broken models}
\noindent
                       mass scale squared. For this reason the
MSSM requires an additional discrete symmetry called matter
parity to forbid them. Under this symmetry the quark and lepton
superfields appearing in the superpotential change sign while the
Higgs superfields are left invariant. Thus the      four  terms
of eq.\rsym\              change sign under this symmetry and are
forbidden while the terms of eq.\superp\           are invariant
and allowed. Using this symmetry the allowed couplings of the
MSSM are only those of eq.\superp           .

Note that supersymmetric states only occur in pairs in
e.g., eq.\yuki\ ,
giving rise to a conserved `` R-parity'' under which the standard
model states are R even and the new supersymmetric states are R
odd. (R-parity is broken if any of the terms of
eq.\rsym \             are present). An immediate consequence of
R-parity is that the new
supersymmetric states may only be produced in pairs and the
lightest supersymmetric state (the LSP) is stable.

     However it is not necessary to forbid all the terms of
eq.\rsym\ to stabilise the nucleon \HALL ,\ZWI \ . The graph of
Fig.1  is the dominant one responsible  for nucleon decay and it may
be seen from it that (redefining fields so that $\mu {'}_i=0$),
provided at least the second or third                      baryon- or
lepton- number violating vertices of eq.\rsym \             is
absent, the graph           vanishes  giving an acceptable model
in which the remaining
operators will violate matter- and R- parity. For example the
first term gives rise to the Yukawa couplings

$$
\lambda_{i j k} ( L_{i} L_{j} {\tilde{e}}_{k} + L_{i}
{\tilde{L}}_{j}     {E}_{k} + {\tilde{L}}_{i} L_{j}     {E}_{k}
)
\eqn\rsyuk
$$
involving a coupling to {\it single} sparticle states. The effect
of these terms is to significantly change the phenomenology
associated with the new supersymmetric states for the LSP may
decay changing $E_{T}$ missing signals to visible energy
 \RPB ,\RP ,\DR\
so it is important to ask whether such models are reasonable. This
amounts to a discussion whether a discrete symmetry can lead to
a non-zero subset of the terms of eq.\rsym    . Indeed this is the
case and a general classification of discrete symmetries
forbidding or allowing different couplings has been recently
worked out  \DGSB .

The main difference between the discrete symmetry leading to the
MSSM and those leading to R-parity breaking schemes is that in
the latter the quarks and leptons transform differently. Is this
to be expected? The origin of discrete symmetries must lie in the
underlying (Grand) unified theory. In string theories it is known
that discrete symmetries arise on compactification under which
quarks and leptons {\it do} transform differently. Similarly many
Unified theories do have gauge symmetries under which quarks and
leptons transform differently and, on spontaneous breakdown, may
lead to such discrete symmetries. A survey of     $Z_{N}$
(flavour-independent) symmetries (for low $N$)
                                 \DGSB\ shows it is easy to obtain any
of the allowed possibilities which forbid fast proton decay, namely
$$
\eqalign{
      Matter\  parity \ \ \   \tilde{\lambda}&=\tilde{\lambda'}=
\tilde{\lambda''}=0 ;\ \ \   \Delta B=\Delta L=0      \cr
      Lepton\  ``parity"\ \ \    \tilde{\lambda'}&= \tilde{\lambda''}=0 ;
\ \ \  \Delta B\not=0,\Delta L=0                     \cr
      Baryon \ ``parity"\ \ \  \tilde{\lambda}&=0 ;\ \  \Delta B=0,\Delta
L\not=0             \cr }
$$
Overall there are many possibilities (including flavour quantum
numbers there are 45 distinct new operators in
eq.\rsym\   !         ).  Although constrained by their virtual
corrections to standard model processes \RPB\       and by
baryogenesis \CAMP\       their coefficients may be large enough
to give rise to completely new SUSY phenomenology.

It has been suggested \WK\
          that the possible discrete symmetries should be further
constrained by the requirement that they be stable against large
gravitational corrections \DGSA ,\DGSB\ and this is only possible
if the symmetries come from an underlying gauge symmetry.
If this is the case, the discrete symmetries have to obey certain
``discrete anomaly'' cancellation conditions \DGSA  . One then finds that
                       the condition for anomaly cancellation in
the underlying gauge symmetry can only be satisfied by a
restricted set of discrete symmetries if one demands only the
minimal light spectrum of states. In this case there are only two
preferred discrete symmetries \DGSB      , $Z_{N}$, for $N<   9$,
namely the usual $Z_{2}$ matter parity leading to the MSSM and
a $Z_{3}$ baryon ``parity" under which
$$
\eqalign{
g(Q,U,D,L,E)&=(1,\alpha^{2},\alpha,\alpha^{2},\alpha^{2})  \cr
\psi_{i} \rightarrow g_{i} \psi_{i}\ &;\
      \alpha=exp(i {{2\pi }\over 3})   \cr }
\eqn\bpar
$$
The baryon parity has the advantage that it
                  additionally forbids the dangerous dimension
5 QQQL operators which may give fast proton decay. At the very
least it should be considered on a par with the MSSM as a minimal
supersymmetric extension of the standard model. One can also find
a $Z_3$ anomaly-free lepton parity but it requires that the
underlying theory posseses a large     $Z_{m9}, m\in {\bf Z}^{+}$
symmetry which looks  rather unlikely \DGSB .

In what follows we will be mainly concerned with the MSSM as a
minimal supersymmetric extension of the standard model.
In fact the mechanism of electroweak breaking does not really
depend on which symmetry one is imposing  to the Yukawa
couplings and practically all the results obtained below
do also apply to the R-parity violating models.

\bigskip

\leftline{\bf 2 Supersymmetry breaking and the MSSM sparticle masses.}

Supersymmetry must be broken in a realistic theory for the
suppersymmetric partners have not been observed  so far. However
the scale of supersymmetry breaking and the associated masses of
the supersymmetric states cannot be far above the electroweak
breaking scale otherwise the hierarchy problem will reappear and
radiative corrections will drive up the electroweak breaking
scale.

Considerable effort has been devoted to understanding the origin
of supersymmetry breaking. If no mass scale is to be added by
hand to the theory just to trigger the breaking then the scale
of supersymmetry breaking must be related to the underlying
scales in the theory, the Planck scale or the unification scale.
In this case it is necessary to understand why the supersymmetry
breaking scale is so small. It is known that if supersymmetry is
broken through quantum   corrections it must be through non-
perturbative effects. This leads to the hope that the large
difference between the Planck scale and the supersymmetry
breaking scale may be understood due to the appearance of a small
factor, $exp(-a/g^2)$,         associated with such non-
perturbative effects. (Here a is a constant , g is a coupling and
the expansion of the exponential around g=0 vanishes,
characteristic of non-perturbative processes.) In our opinion
this is the only promising explanation that has been advanced to
explain the large mass hierarchy and so we concentrate on the
implications for the spectrum of supersymmetric states that
results from models implementing this idea.

Nonperturbative effects occur when an interaction become large
and in supersymmetric gauge theories there is a very natural
source of such effects leading to supersymmetry breaking. This
follows because an asymptotically free (non-abelian) gauge
interaction, initially small, increases in strength as the energy
scale {\it decreases} and at some scale will become large and in
the non-perturbative domain. In analogy with QCD it may be
expected that the gauginos will be strongly bound by this
interaction to form a gaugino condensate, $<\lambda \lambda>$
which breaks (local \CFGVP\ )
                     supersymmetry generating a gravitino mass
$m_{3/2}        \propto     {{<\lambda \lambda>}\over  {M_{P}^{2}}}  $
\FGN ,\GAUGINO ,\DIN ,\CLMR ,\FILQ ,\DUAGAU .
Since the scale determining the gaugino condensate is given by
the scale, $\Lambda$, at which the gaugino binding becomes non-
perturbative we have
$$
                      m_{3/2}        \propto     {{\Lambda^{3}}\over
{M_{P}^{2}}}= M_{P}\ e^{({ 3\over { 2b_{0}g^{2}}})}
\eqn\ssb
$$
where $b_0$ (which is assumed to be negative) is
the coefficient of the one loop $\beta$-function and the last
equality follows from using the running of the gauge coupling
from its value, g, at the Planck scale. This has the form
anticipated in our discussion of non-perturbative effects and,
for suitable values of g and $b_{0}$, offers an explanation for
the large magnitude of the mass hierarchy.

In order to build a realistic theory using this mechanism it is
usual to associate the gaugino condensate with an ``hidden''
sector which couples to the visible sector only via gravitational
interactions. This assumption is convenient in order to
obtain universal supersymmetry-breaking effects \IBA       ,\BFS .
It is also the natural situation in 4-D strings  \GAUGINO ,
\DIN , \CLMR ,\FILQ ,\DUAGAU  .
in which the gauge group is always larger than the one of the
standard model or its GUT extensions. Very often there are in these
theories ``hidden sectors'' of particles which couple to hidden sector
(confining) gauge interactions
which couple only gravitationally to the ``observable'' world.
                                                Hidden sector
breaking      has the advantage of reducing the effect of
supersymmetry breaking in the visible sector because the
gravitational strength couplings generate visible sector masses
of order the gravitino mass and not of order the much larger
gaugino condensate scale. Apart from triggering supersymmetry
breaking the hidden sector plays no role in low energy
phenomenology for the states of this sector are confined with
mass of order the gaugino condensate scale ($\sim
10^{13}Gev$).

The  most important implication for the MSSM
                           of such hidden sector breaking is that
supersymmetry breaking is communicated to the visible sector only
via gravitational interactions. The precise effect on the visible
sector depends on the details of these gravitational interactions
but the independence of these interactions from the gauge and
Yukawa interactions of the MSSM leads to a useful
parameterisation of the supersymmetry breaking effects in terms
of flavour symmetric breaking terms \IBA ,\BFS ,\HLW . This amounts to
including a common mass $m_{1/2}$            for the gauginos and another
common mass, $m_{o}$, for the scalars. The gauge bosons and
fermions do not acquire mass at this stage due to residual
unbroken gauge and chiral symmetries. It is also found in
specific supergravity models that there are additional \OW ,\IBA ,
\BFS\ supersymmetry breaking terms given by ($A_{0}P_{3}$ +
$B_{0}P_{2}$) where $A_{0}$ and $B_{0}$ are masses of order
$m_{3/2}$         and $P_{3}$ and $P_{2}$ are the trilinear and
quadratic terms of the superpotential with the supermultiplets
replaced by their scalar components. The expectation for $A_{0}$
and $B_{0}$ depends on the metric \HLW .

The flavour independence of the supersymmetry breaking terms is
broken by radiative corrections \IRB\ involving the gauge and Yukawa
couplings of the standard model. These corrections may be
calculated explicitly and are most conveniently included via the
renormalisation group equations \INOUB ,\IBA ,\AGPW ,\EHNT\ for the
masses (cf Appendix)
and the A and B parameters using for initial values at $M_{X}$
the common gaugino and scalar masses $m_{1/2}$         and
$m_{o}$ and the common $A_{0}$ and $B_{0}$ parameters. Note that
even if some of these initial values are zero the related terms
may be generated at low scales via radiative
corrections.\footnote*{Recently \IL\  it has been observed that in
string theories the universality of supersymmetry breaking mass
terms may be broken if the fields have different modular weights.
Although possible the necessity to avoid large flavour changing
neutral currents strongly constrains the amount of such non-
universality and suggests that, in a viable model, flavour blind
scalar masses at the unification scale is a good approximation.
On the other hand, non-universal gaugino masses may also be
pressent in certain string models \IL\  .}

 The final form for the effective potential from the SUSY
breaking soft terms for one family and keeping only the up quark
Yukawa coupling is

$$\eqalign{
V_{eff}\  &=\  [hA({\tilde Q}{\tilde U} H_{2}      )+h.c. ]  +
 [B\mu (H_{1}    H_{2}                       )+h.c.] \cr
 &\mu _1^2   \mid H_1 \mid ^2
+\mu _2^2   \mid H_2 \mid ^2
+m_{{\tilde L}}^{2}                 \mid {\tilde L} \mid^{2}
+m_{{\tilde E}   } ^{2}\mid {\tilde E }   \mid^{2}   \cr
& +m_{{\tilde Q}}^{2} \mid {\tilde Q} \mid^{2}
+m_{{\tilde U}    }^{2}\mid {\tilde U}    \mid^{2}
+m_{{\tilde D}    }^{2}\mid {\tilde D }   \mid^{2}   \cr }
\eqn\soft
$$
At the unification (Planck) mass one assumes
$$\eqalign{   m_{\tilde Q}^2=m_{\tilde U}^2=m_{\tilde D}^2
      &=m_{\tilde L}^2=m_{\tilde E}^2  \ =           \ m_0^2  \cr
\mu _1^2 =\mu _2^2  \ &=\ m_0^2\ +\ \mu _0^2 \cr }
\eqn\masboundary
$$
where $\mu _0$ is the boundary value for $\mu $.
One also has universal gaugino masses
$$
M_1=M_2=M_3\ =\ m_{1/2}   \ .
\eqn\magaugi
$$
Using this parameterisation for the soft supersymmetry breaking
mass terms at the unification scale it is now straightforward to
determine the spectrum of the light states of the MSSM, for the
couplings and soft supersymmetry breaking terms at low energy
scales may be determined in terms of their values at the
unification scale using the coupled renormalisation group
equations to determine the radiative corrections. We turn now to
a discussion of these equations and their implications for the
generation of a further stage of spontaneous symmetry breaking
this time in the gauge sector.

\bigskip
\leftline{\bf 3  Electroweak breaking}
\bigskip
We discussed   in the previous section the possible origin of the
required breaking of supersymmetry. The resulting Lagrangian
includes supersymmetry-conserving pieces
plus extra terms breaking supersymmetry explicitly (but softly \GIRAR ).
The next problem to address is the breaking of the
weak interactions symmetry $SU(2)\times U(1)$. In principle,
this breaking is completely independent of that of supersymmetry
but the fact that the masses of sparticles have to be of order
the weak scale makes one suspect that both symmetry-breaking
processes should be somehow related. Indeed, as we will now
describe, once one has broken supersymmetry the breaking of
$SU(2)\times U(1)$ appears as an automatic consequence of
quantum corrections.

To study the process of electroweak symmetry breaking let us
consider the piece of the scalar potential involving just the
Higgs doublets. From the gauge interactions and the interactions
following from the superpotential of           eq(4) one has

$$ \eqalign{
V(H_1,H_2)\ &=\  {{{g_2}^2}\over 2}\ (H_1^{*}{{\tau ^a}\over
2}H_1+
   H_2^{*}{{\tau ^a}\over 2}H_2)^2\ +\ {{{g_1}^2}\over 8}\
(|H_1|^2-|H_1|^2)^2\  \cr
&+\ {\mu _1}^2\ |H_1|^2\ +\ {\mu _2}^2\ |H_2|^2\ -\ {\mu _3}^2
(H_1H_2\ +h.c.\ )  \cr }
\eqn \pot
$$
where $\tau ^a,a=1,2,3$ are the $SU(2)_L$ Pauli matrices and
$$
{\mu _1}^2\ \equiv m_0^2\ +\ \mu _0^2\ ;\ {\mu _2}^2\ \equiv \
m_0^2\ +
\ \mu _0^2\ ;\ {\mu _3}^2\ \equiv \ B_0\mu _0  \ .
\eqn \bounda
$$
This is the SUSY version of the `mexican hat' Higgs potential of
the standard model. However, this potential as it stands looks
problematic. Indeed, in order to get a non-trivial minimum
we need to have a negative $\it mass^2$ eigenvalue in the Higgs
mass matrix, i.e., we need $\mu _1^2\mu _2^2-\mu _3^4\ <0$.
However, since $\mu _1^2=\mu _2^2>0$, this may only happen if
$\mu _1^4=\mu _2^4<\mu _3^4$ in which case the scalar
potential is unbounded below in the direction
$<H_1>=<H_2>\rightarrow \infty $.

\bigskip

\leftline{\bf 3.1 One loop radiatively corrected potential}
\bigskip

The puzzle is resolved \IRB\  by noting that the boundary
conditions
eq.\bounda\   apply only at the unification or Planck scale.
At any scale below one has to consider the quantum corrections
to the scalar potential which can be substantial. Consider
for example the one-loop corrections to the masses of the
Higgs fields. There are graphs as in fig.2

\vskip 4.truein

\centerline{\bf Fig. 2}
\noindent
which involve the
Yukawa couplings of the Higgs field $H_2$ to the u-type
quarks and squarks. Of course, these corrections will be
negligible except for the ones involving the top quark which
has a relatively large Yukawa coupling
 (for simplicity we ignore here the possibility of a large
bottom Yukawa coupling).
While supersymmetry is
a good symmetry the first graph in fig. 1 leads to a (negative)
quadratically divergent contribution which is exactly cancelled
by the second graph. Once susy is broken the sparticles get
masses and the right-hand diagram is suppressed compared to the
left-hand one leaving an overall uncancelled $negative$ contribution \IRB
$$
\delta \mu _2^2\ \simeq \ -{3\over {16\pi ^2}}h_t^2m_{\tilde Q}^2
\log (M_X^2/m_{\tilde Q}^2)\ .
\eqn \semin
$$
If $h_t$ is large enough (i.e., if the top quark is heavy
enough) this negative contribution may overwhelm the original
positive contribution and trigger electroweak symmetry breaking.
Similar diagrams exist for the other Higgs field $H_1$ but
those are expected to give small contributions since they
will be proportional to the bottom Yukawa coupling.

One may worry that one can find similar graphs involving squarks
such as ${\tilde t},{\tilde b}$ that could drive their $\it
mass^2$  negative leading to minima with broken charge and
colour. Indeed these
graphs exist but coloured scalars also get large ({\it positive})
contributions to their $\it mass^2$ from loops involving
gluinos which are proportional to the large strong coupling
constant, preventing $SU(3)\otimes U(1)_{em}$ breaking.  Thus one
sees that the structure of the
minimal supersymmetric standard model is such that quantum
corrections select the desired pattern of $SU(3)\otimes
SU(2)\otimes U(1)$
symmetry breaking in a natural and elegant way.

With $\mu _2 ^2\not= \mu _1^2$ the potential in  eq.\pot  is
perfectly well behaved and one can see it is minimized for \INOU
$$
\nu ^2\ \equiv \ \nu _1^2+\nu _2^2\ =\ {
{2(\mu _1^2-\mu _2^2-(\mu _1^2+\mu _2^2)cos2{\beta })} \over
{(g_2^2+g_1^2)cos 2\beta } }
\eqn \min
$$
where $\nu _{1,2}=<H^0_{1,2}>$ and $sin2\beta \equiv 2\mu _3^2/
(\mu _1^2+\mu _2^2)$.
The existence of a non-vanishing $\mu _3^2$ forces the two vevs
to be aligned in such a way that electric charge remains unbroken.
The W mass is related to $\nu$ via
$\nu^2=2M_W^2/g^2$. This condition may be equivalently
written \IBLO ,\ILM
$$
{{\nu _2^2}\over {\nu _1^2}}\ =\  {{\mu _1^2+{1\over 2}M_Z^2}
\over {\mu _2^2+{1\over 2}M_Z^2}}  \
\eqn \const
$$
where $\mu ^2_{1,2}$ should be evaluated at the weak scale.
Thus in a model with the correct $SU(2)\times U(1)$ breaking  the
parameters are constrained in such a way
that $\mu ^2_{1,2}(M_W)$ and $\mu ^2_3(M_W)$ satisfy the above
conditions. As we discussed in section(2) the free parameters in
the minimal model are
just
$$  m_0\ ,\ m_{1/2}\ ,\ A_0\ ,\ \mu ^2_{03}\equiv B_0\mu _0\ , \ \mu _0
\eqn \param
$$
plus the Yukawa couplings, of which      $h_t$       is likely
to be    the only one
playing an important role in the running of the soft terms. In
order
to see how all these parameters are constrained we need to
use the renormalization group equations which relate the values
of couplings and masses at the unification scale with their
values at the weak scale. The renormalisation group equations
generalise  the analysis of radiative corrections given in
eq.\semin\    allowing for the summation of all powers of the
logarithmic corrections. For completeness the renormalization
group
equations are provided                  in the appendix.
Semi-analytic solutions to those equations may be found in ref.\IBLO
,\ILM .

Before describing the renormalization group running  let us
make a few comments on the general analysis presented  here.
First, one should not keep the evolution determined from a
renormalization group
equation below the threshold of any particle involved in the
given equation \footnote*{As may be seen from eq.\semin\ this is
necessary to reproduce the true result obtained from the evaluation
of the Feynman graphs on shell.}.
                In other words, care has to be taken with the
thresholds of the different sparticles. Usually it is enough to
use a step function for each threshold. These threshold effects
are discussed e.g. in \REFIN .
Secondly, for a sufficiently large $A$ parameter, there
can appear other minima in the scalar potential favourably
compeating  with that in eq.\min\ . They involve vacuum expectation
values for sleptons and/or squarks and they break charge conservation.
To avoid that it is enough to constraint               $\mid A\mid $
to sufficiently small   values (typically $\mid A\mid \leq 3$
\AAA ,\AGPW ,\IBLO ,\AAB .)
Thirdly, similar results to those presented below are obtained in
the case with an additional singlet $N$ as discussed in section 1,
eq.\singli .  For small $\lambda '$ the analysis of the
electroweak symmetry breaking \DS ,\IM \ is rather  similar although the
existence of a term $\mid H_1H_2\mid ^2$ in the scalar potential
allows for symmetry breaking only for $m_t\geq 70$ GeV
(for a numerical analysis of this case see \IM ,\DREE\ .)
Finally, in the discussion below we have considered the most probable
case in which $h_t$ is assumed to be much bigger than $h_b$.
Numerical studies including a non-negligible $h_b$  were
presented  in \IBLO ,\DREEB .
We will not describe all these fine details here but
merely give a general discussion of the most prominent
physical implications and direct the reader to those references
for further details.
\bigskip

\leftline{\bf 3.2 Renormalisation Group analysis}
\bigskip

The most extensively studied renormalization group equations
are those concerning the three gauge coupling constants. The
extrapolation of the measured values of $\alpha _{e.m.}$ and
$\alpha _3$ to high energies shows that for the MSSM the couplings
meet for a value of the
Glashow-Weinberg angle $sin^2\theta _W\simeq 0.23$ \DG ,\EINHJON\ in
extremely good agreement with recent LEP data \AMALDI ,\RR  . We discuss
this point in some detail in the next section. The
renormalization group equations for the Yukawa couplings \INOU\
(see the appendix)   are  also of considerable
           interest.    They can be integrated analitically in the case
in which one only keeps the top-quark Yukawa coupling. For the
third generation one finds  \IBLO
$$
\eqalign{h_t^2(t)\ &=\ h_t^2(0)\ {{E_1(t)}\over
{1+6Y_t(0)F_1(t)}}  \cr
h_b^2(t)\ &=\ h_b^2(0)\ {{E_2(t)}\over {(1+6Y_t(0)F_1(t))^{1/6}}}
\cr
h_{\tau }^2(t)\ &=\ h_{\tau }^2(0)\ E_3(t) \ , \cr }
\eqn \renyuk
$$
where $t\equiv 2\log (M_X/Q)$ and $E_{1,2,3}$ and $F_1$ are known
functions given in the appendix and $Q$ is the scale at which the
couplings are evaluated. The $E_i$ functions give just the usual
gauge anomalous dimension enhancement whereas the effect of the
top Yukawa coupling in the running gives the extra factor.  Let
us first discuss the case of the top quark. Notice that for small
$h_t(0)$ one recovers the well-known gauge anomalous dimension
result. However, for $Y_t(0)\rightarrow \infty$ one gets
$$
h_t^2(t)\ =\ {{(4\pi )^2E_1(t)}\over {6F_1(t)}}
\eqn \htop
$$
independently of the original value of $Y_t(0)$, i.e., there is
an infrared fixed point. At the weak scale       ($t\simeq 67$) one
obtains $E_1\simeq 13$ and $F_1\simeq 290$ which gives an upper
bound  for the top-quark mass
$$
m_t\ =\ h_t\nu _2\  \leq \ h_t\nu\ \leq \ 190  \ GeV \ .
$$
One also observes in eq. \renyuk\  that the bottom-quark mass
decreases as $m_t$ increases \INOU ,\IBLO whereas $m_{\tau }$ does not
depend directly on $m_t$ (see below).
\vskip 4.truein
\centerline{\bf Fig. 3-a}
\endpage
Let us now consider the running of the mass parameters which are
the ones of direct relevance to the $SU(2)\times U(1)$-breaking
process. In particular, consider the running of the squarks,
sleptons and Higgs masses (see                          the
appendix).
\vskip 4.truein
\centerline{\bf Fig. 3-b}
           The renormalization group equations describing the mass$^2$
evolution       have a gauge contribution proportional
to gaugino masses and a second contribution proportional to the
top-Yukawa coupling$^2$. The gauge piece makes the mass$^2$  increase
as the energy decreases. In particular, squarks get more and more
massive as we go to low energies since their equation is
proportional to $\alpha _3$. The piece in the equations
proportional to the top
Yukawa coupling has the opposite effect and decreases the
mass$^2$ as the scale decreases. This effect is normally not big
enough to overwhelm the large positive contribution to the
mass$^2$ of squarks involving the QCD coupling but may be
sufficiently large to
overwhelm the positive contribution of weakly interacting scalars
which only involve the electroweak couplings.
The only such scalar in which this can happen is $H_2$ since it
is the only one (unlike sleptons) which couples directly to
the top Yukawa.  This is nothing but the renormalization group
improved version \IBA ,\IBLO ,\AGPW ,\EHNT\
                 of the mechanism in eq.\semin\  . We thus see
that the quantum structure of the MSSM leads automatically
to the desired pattern of symmetry breaking in a natural way.
The qualitative behaviour of the running of scalars is shown in
fig. 3. A quantitative analysis of the process was given in \IBLO ,
\AGPW ,\EHNT ,\JRR ,\ILM .

Apart of obtaining the desired pattern of symmetry breaking
one is interested in finding out the spectrum of
sparticles in this scheme. Since there are only a
few free parameters one has  strong predictive power. In the case of
the squarks and sleptons integration of the renormalization
group equations (neglecting Yukawa couplings) leads to the
following result \IBLO
$$
\eqalign{
 m^2_{{\tilde U}_L}\ &=\ m_0^2+2m_{1/2}^2({4\over 3}{\tilde {\alpha
}}_3f_3
+{3\over 4}{\tilde {\alpha }}_2f_2+{1\over {36}}{\tilde {\alpha
}}_1f_1
)+cos(2\beta )M_Z^2({{-1}\over 2}+{2\over 3}sin^2\theta _W)  \cr
 m^2_{{\tilde D}_L}\ &=\ m_0^2+2m_{1/2}^2({4\over 3}{\tilde {\alpha
}}_3f_3
+{3\over 4}{\tilde {\alpha }}_2f_2+{1\over {36}}{\tilde {\alpha
}}_1f_1
)+cos(2\beta )M_Z^2({1\over 2}-{1\over 3}sin^2\theta _W)    \cr
 m^2_{{\tilde U}_R}\ &=\ m_0^2+2m_{1/2}^2({4\over 3}{\tilde {\alpha
}}_3f_3
+{4\over 9}{\tilde {\alpha }}_1f_1)-cos(2\beta )M_Z^2({2\over 3}
sin^2\theta _W)   \cr
 m^2_{{\tilde D}_R}\ &=\ m_0^2+2m_{1/2}^2({4\over 3}{\tilde {\alpha
}}_3f_3
+{1\over 9}{\tilde {\alpha }}_1)+cos(2\beta )M_Z^2({1\over 3}
sin^2\theta _W)    \cr
 m^2_{{\tilde E}_L}\ &=\ m_0^2+2m_{1/2}^2({3\over 4}{\tilde {\alpha
}}_2f_2
+{1\over 4}{\tilde {\alpha }}_1f_1)+cos(2\beta )M_Z^2({1\over 2}-
sin^2\theta _W)    \cr
 m^2_{{\tilde \nu }_L}\ &=\ m_0^2+2m_{1/2}^2({3\over 4}{\tilde {\alpha
}}_2f_2
+{1\over 4}{\tilde {\alpha }}_1f_1)-cos(2\beta ){1\over 2}M_Z^2
\cr
 m^2_{{\tilde E}_R}\ &=\ m_0^2+2m_{1/2}^2({\tilde {\alpha }}_1f_1)+
cos(2\beta )M_Z^2sin^2\theta _W             \cr }
\eqn \masillas
$$
where $\theta _W$ is the weak angle, $M_Z$ is the $Z^0$ mass
and $m_0,m_{1/2}$ and $tg\beta =\nu _2/\nu _1$ are
related to the free parameters in
 eq.\param .
In this equation
$$
{\tilde {\alpha }}_i\ \equiv \ {{\alpha _i(M_X)}\over {(4\pi )}}\
;\
f_i\ \equiv \ {{ (2 +b_i{\tilde {\alpha }}_it)}\over
{(1+b_i{\tilde {\alpha }}_it)^2}}t
\eqn \auxil
$$
where $b_i=(-3,1,11)$ are the one-loop coefficients of the $\beta
$-function of
the $SU(3)\otimes SU(2)\otimes U(1)$ interactions. The above
equations
assume universal soft masses $m_0$ for all the scalars in the
theory
at the unification scale as well as universal gaugino masses $m_{1/2}$.
The rightmost term in eqs.\masillas \ does not in fact come from
the
integration of the r.g.e.'s but from the contribution of the
D$^2$-term in the scalar potential of sfermions once $SU(2)\times
U(1)$ is broken. As noted above  the squarks will be heavier than
the
sleptons since $\alpha _3\gg \alpha _2$.

When computing physical masses the scale $Q$ should be chosen equal
to the mass. Fot masses in the TeV scale this corresponds to
$t\simeq 67$. With this value the coefficients of the
                    $m_{1/2}^2$-terms  in  eq.\masillas
                       are approximately $7.6$ for ${\tilde U}_L,
{\tilde D}_L$, $7.1$ for ${\tilde U}_R,{\tilde D}_R$,
$0.53$ for ${\tilde E}_L,{\tilde \nu}_L$ and $0.15$ for
${\tilde E}_R$. The formulae for the stop and sbottom are
modified by the top Yukawa coupling effects which can be evaluated
numerically. Furthermore there are two additional contributions
to the scalar masses: one is just equal to the mass of the
corresponding fermionic partner $m_f$ and the other mixes left and
right-handed sfermions and is proportional to
$m_f A$. These two contributions are only relevant for the
heaviest quarks and may be included in a numerical analysis
as discussed in section 4.

Let us also briefly discuss the masses of physical Higgs particles.
The MSSM contains two doublets of Higgses $H_1=(H_1^{+},H_2^{0}),
H_2=(H_2^{-},H_2^{0})$ with            eight degrees of freedom.
Three         are swallowed by the $W$s and the $Z^0$ as they become
massive. There remain one charged Higgs scalar $H^{\pm }$,
two neutral scalars $h,H$ and one neutral pseudoscalar $P$. Using the
scalar potential \pot\ one finds for their masses  \INOU
$$
\eqalign{
m^2_{H^{\pm }}\ &=\ M_W^2\ +\ \mu _1^2\ +\ \mu _2^2   \cr
m^2_{P}\ &=\ \mu _1^2\ +\ \mu _2^2      \cr
m^2_{H,h}\ &=\ {1\over 2}(m^2_P+M_Z^2\pm ((m^2_P+M_Z^2)^2
-4m^2_PM_Z^2cos^22\beta )^{1/2})   \cr }
\eqn \higgs
$$
Using the above expressions one obtains the following constraints
on Higgs masses  \INOU ,\FLSH
$$
\eqalign{
0\ &\leq \ m_h\ \leq \mid cos2\beta \mid M_Z    \cr
m_h\ &\leq \ m_P\ \leq \ m_H    \cr
m_H\ &\geq \ M_Z   \cr
m_{H^{\pm }}\ &\geq \ M_W  \cr
} \eqn \ineq
$$
so that one observes that the scalar $h$ is always necessarily
lighter than the $Z^0$. Although this is an important constraint,
one must emphasize that for large $m_t$ these tree level
Higgs formulae get important loop corrections which may modify
the above inequalities \HAHE .

Let us now briefly discuss the spectrum of the fermionic sparticles.
For the case of the gluino it is very easy to relate its mass
to the SUSY-breaking parameters. It is simply given by
$$M_{\tilde g}=({{\alpha _3}\over {\alpha _{GUT}}})\ M.\eqn \glue $$

    The mass spectrum for the ${\tilde W},{\tilde B}$ and Higgsinos
is more complicated because once $SU(2)\times U(1)$ is broken
they mix amongst themselves.
                                     Apart from this mixing,     there
are the usual Majorana mass for the gauginos             and the
direct Higgsino mass $\mu $ coming from the superpotential. In this way,
we have a mass matrix for the charged winos-Higgsinos:
$$\left( {\matrix{{\tilde W}^- & {\tilde H}^- \cr }} \right)
\left( {\matrix{M_2 & g_2{\nu _2} \cr
g_2{\nu _1} & \mu \cr }}\ \  \right)
\left( {\matrix{ {\tilde W}^+ \cr {\tilde H}^+ \cr }} \right)
\eqn \charginos $$
where $M_2$ is the direct wino mass and ${\nu _{1,2}}=<H_{1,2}>$.
This matrix has two eigenstates ${{\tilde {\chi }}_1}^{\pm }$
and ${{\tilde {\chi }}_2}^{\pm }$
(the `charginos')
with mass eigenvalues
$$\eqalign{ {M_{2,1}}^2\ &= {1\over 2}({M_2}^2+{\mu }^2+2{M_W}^2  \cr
&\pm{1\over 2}\sqrt{({M_2}^2-{\mu }^2)^2+4{M_W}^4cos^22\beta +
4{M_W}^2({M_2}^2+{\mu }^2+2M_2\mu sin2\beta )) .}   \cr}
\eqn \chargmass $$ The lightest of them,
${{\tilde {\chi }}_1}^{\pm }$, is very often lighter than $M_W$ and
has a good chance of being the lightest charged SUSY particle.

The mass matrix for the `neutralinos' is more complicated. In a basis
spanned by $({\tilde W}^0,{\tilde B}^0,{{\tilde H}_1}^0,
{{\tilde H}_2}^0)$ one finds \NILL \ :
$$M_{{\chi }}^0\ =\ \left( {\matrix{
M_2 & 0 & {{-g_2{\nu _1}}\over {\sqrt{2}}} &
{{g_2{\nu _2}}\over {\sqrt{2}}} \cr
0 & M_1 & {{g_1{\nu _1}}\over {\sqrt{2}}} &
{{-g_1{\nu _2}}\over {\sqrt{2}}} \cr
{{-g_2{\nu _1}}\over {\sqrt{2}}} &
{{ g_1{\nu _1}}\over {\sqrt{2}}} & 0 & \mu \cr
{{ g_2{\nu _2}}\over {\sqrt{2}}} &
{{-g_1{\nu _2}}\over {\sqrt{2}}} & \mu & 0 \cr}} \right)
\eqn \neutralinos $$
where $M_2$ and $M_1$ are related through the renormalization group
equations by       $M_1=(3{\alpha _1}/5{\alpha _2})M_2$.

This matrix has four eigenvalues and four corresponding
eigenstates ${{\tilde {\chi }}_i}^0$ ($i=1-4$), the
 neutralinos.  The lightest of them
(say ${{\tilde {\chi }}_1}^0$) has good chances of being the
lightest supersymmetric particle (LSP). It will be a linear combination
of the original fields:
$${{\tilde {\chi }}_1}^0\ =\
U_{11}{\tilde W}^0+U_{12}{\tilde B}+U_{13}{{\tilde H}_1}^0+
U_{14}{{\tilde H}_2}^0  \eqn  \lsp $$
where the unitary matrix $U_{ij}$ relates the ${{\tilde {\chi }}_i}^0$
fields to the original ones. The entries of that matrix will depend
only on $tg\beta $,$M_2$ and $\mu $ and, depending on those
parameters, ${{\tilde {\chi }}_1}^0$ will be more
`Higgsino-like' or `photino-like' etc. For example,
for $U_{13}=U_{14}=0$ the LSP would be the photino
${\tilde {\gamma }}=sin\theta _W{\tilde B}+cos\theta {\tilde W}^0$.
In general one has to diagonalize the matrix \neutralinos\ \
case by case in order to identify the eigenstates and mixing
angles of the neutralinos. Notice that the couplings of
neutralinos to the other particles will thus be affected
by mixing angles. Examples of SUSY spectra consistent with
appropriate electroweak symmetry breaking are provided in
table 2.

 Finally we comment on the renormalisation group improvement of
Grand unified (GUT) predictions \BEGN\ for quark and lepton masses. For
example, in SU(5) $m_{b}$ and $m_{\tau}$ are equal at the GUT
scale. The gauge corrections to that relationship in the supersymmetric
case \EINHJON\  are numerically similar to the non-SUSY case.
 However, for a heavy top quark using eq.\renyuk\
one gets the result          \IBLO\
$$
{{m_b(t)}\over {m_{\tau }(t)}}\ =\ ({{\alpha _3(t)}\over
{\alpha _3(0)}})^{8/9}({{\alpha _1(t)}\over {\alpha
_1(0)}})^{10/99}
(1\ +\ 6Y_t(0)F_1(t))^{-1/12} \ .
\eqn \mbmt
$$
Of course, below the supersymmetric and top thresholds one has
to
use the equivalent non-supersymmetric (top-less) equations.
This equation shows that for a sufficiently heavy top-quark the
$m_b/m_{\tau }$ ratio is substantially decreased. This effect
is numerically displayed in fig. 4.

Notice that if one had an
arbitrarily good precision of $\alpha _3$ (and a reliable
definition
of the b-quark current-mass) one should be able to extract a
prediction
\noindent
           for the mass of the top \IBLO . Unfortunately this is not the
case.
More recent analysis of the effect of a large t-quark mass on the
$m_b/m_{\tau }$ ratio may be found in \MBMT .
\endpage
{}.
\vskip 4.truein
\centerline{\bf Fig. 4}

\bigskip

\bigskip

\leftline{\bf 4\ Numerical analysis}
\bigskip

Using the parameterisation for the soft supersymmetry breaking
mass terms at the unification scale given in section 2
together with the renormalisation group equations of
section 3 and the appendix, it is straightforward to determine the
supersymmetry breaking terms at low energy scales. As discussed
in the last section this may lead automatically to a stage of
electroweak breaking allowing for the complete determination of
the mass spectrum of the states of the MSSM, including squarks,
sleptons and the W and Z masses, in terms of the couplings of the
theory and the parameters determining the supersymmetry breaking
terms\RR . In this section we will generalize the discussion of
section 3 to include all the terms in the RGEs by numerical
integration and show that the resulting pattern
is in excellent agreement with the precision measurements of the
parameters of the standard model. We will also show that the MSSM
leads to predictions for the running gauge couplings in excellent
agreement with a minimal unification picture in which these
couplings are related at a high unification scale. While this
latter result is not a necessary condition for the viability of
the MSSM (unification could be non-minimal with several stages
at high scale) its success does provide some circumstantial
evidence for the need of low-energy supersymmetry. Perhaps more
significant will be the very strong constraint on the
supersymmetry breaking scale that results merely from the
condition that there is a large scale of unification. As we will
discuss, this is insensitive to the details of the unification
and offers the most sensitive test of the idea there should be
a low energy supersymmetry to solve the hierarchy problem, for
it provides a strong upper bound to the masses of the new
supersymmetric states. Although unification is not an absolute
necessity for supersymmetry in our opinion it is the main reason
for preferring the supersymmetric solution to the hierarchy
problem over the composite solution, and for this reason we
consider it important to consider the MSSM within the framework
of an underlying unified theory \SUSYGUT .

\bigskip

\leftline{\bf 4.1 \ Unification of gauge couplings}
\bigskip

The possibility for unification of gauge couplings was noted by
Georgi, Quinn and Weinberg \GQW\ who showed in the standard model that
radiative corrections drive the strong, electromagnetic and weak
couplings together as high energy scales. This may be seen from
the renormalisation group equations for the $SU(3)\otimes
SU(2)\otimes U(1)$ gauge couplings, given in the Appendix.
Those equations   determine  the gauge couplings at a scale
$Q^{2}$ in terms of the initial values at a scale $M_{X}^{2}$
summing the leading and next to leading logarithmic terms in
$log(Q^{2}/M_{X}^{2})$. The initial values are determined by the
Grand Unified theory or compactified
(string) theory at $M_{X}$.  In SU(5)
$\alpha_{i}(M_{X}^{2})=\alpha_{G}$, where
$\alpha_{i}=C_{i}g_i^2/4\pi$             and $C_{2,3}=1,
C_{1}=3/5$.         This also applies to  $SO(10)$ and $E(6)$
GUTs. In four dimensional compactified string theories $C_{i}$
are related to the Kac-Moody level
\endpage
\noindent
of the underlying conformal
field theory \GINS\ with the usual choice of level one leading to the
same values given above.
\vskip 3.truein
\centerline{\bf Fig. 5}
Using these equations it is straightforward to check whether the
low-energy
           couplings evolve in energy to meet at the unification
value $\alpha_{G}$. In Fig.5            the evolution of the
couplings (including two-loop effects)
          is plotted for the non-
\vskip 3.truein
\centerline{\bf Fig. 6}
\endpage
\noindent
   -supersymmetric case,\footnote*{It is convenient to plot inverse
couplings for, cf eq.(37), these are approximately linear in
$\log \mu $.} showing
that the predictions are inconsistent; the couplings fail to meet \AMALDI
\ by more than seven standard deviations (the figures are taken from
Amaldi et al in ref.\AMALDI\ )  .

The situation is quite different when the effect of
supersymmetric states are added when calculating the $\beta$
functions \DG , \EINHJON\ . Including these gives the evolution of
the couplings
                             is shown in Fig. 6           and it
may be seen that the couplings do meet in a point. This happens
if the mass of the new supersymmetric states (assumed degenerate
here) are low $M_{SUSY} \sim      10^{2.5\pm 1}$ GeV.
On the basis
of these results it is tempting to argue that there is evidence
both for new forms of (relatively) light supersymmetric matter
{\it and} an underlying unified theory. The conclusion is so
dramatic it is important to discuss the reliability of the
prediction.

At the two loop order of precision non-logarithmic corrections
due to states with mass of $O(M_{X})$ are also important. These
``threshold'' corrections can be analytically computed \LLR\
and included in the boundary value at the scale $M_{X}^{2}$

$$
\alpha_{i}^{-1}(M_{X}^{2}) = \alpha_{X}^{-1}
+{     1 \over {6\pi}}
{1\over 2} [(t^{H}_{iF})^{2}+4(t^{H}_{iF})^2ln({    {M_{X}}\over {M_
{F}}})+{1\over 2} (t^{H}_{iS})^2ln(    {{M_{X}}\over {M_{S}}})]
\eqn\thres
$$

These terms give the threshold corrections due to massive states;
$M_{X},M_{F}$ and $M_{S}$ are the masses of the massive gauge,
fermion and scalar fields respectively and
$t^{H}_{iV},t^{H}_{iF}$ and $t^{H}_{iS}$ are the matrices which
represent the generators of the gauge group on the superheavy
vector, fermion and scalar fields respectively. In general we
expect a spectrum of heavy fermion and scalar fields in which
case there will be several fermion and scalar terms contributing
to eq.\thres           . Since some of these fields acquire mass
through the Yukawa couplings in the theory     their mass may be
substantially different from $M_{X}$.

These effects were first studied \DA\          in the context of
non-supersymmetric SU(5). Although individual terms are small,
it was found that the totality could be large in non-minimal
versions of the model in which there are scalars transforming as
some high representation of $SU(5)$ giving rise to large numbers
of heavy scalar fields contributing to eq.(2). Recently the same
observation has been made by
Barbieri and Hall \BAHALL\    in the context of supersymmetric
SU(5).

It is clear the possible presence of such massive states at the
unification threshold introduces an inherent
uncertainty in the analysis, as does the assumption of a single
scale of breaking at the unification scale directly to the MSSM
with the standard model gauge group. Due to these uncertainties
it is not possible to make any definitive statements about the
need for unification and a stage of low energy supersymmetry.
Nonetheless it still of interest to study the minimal unification
possibility, for its success may indicate simplicity in the
unification possibly due to the absence of large representations
of heavy states as may happen in some compactified string
schemes, or due to the degeneracy of such states. In any case the
minimal analysis, in which the
predictions are well defined, provides a benchmark to judge
unification predictions and to test the detailed structure of the
low energy supersymmetric theory.

 Apart from the possible threshold effects due to massive GUT
states at $M_{X}$ there may also be corrections in compactified
superstring models of unification due to the tower of Kaluza
Klein and string states with mass quantised in units of the
compactification and string scales \KAPLU ,\DKLB ,\DFKZ ,\AELN ,\ILR ,
\IL .  In such theories there is only one fundamental mass scale,
the Planck scale, and the scale, $M_{X}$, at which the gauge
couplings are related is determined in terms of the Planck scale
by these threshold    corrections
                                     . In a class of
orbifold four dimensional string theories the result of including
these corrections yields boundary values  given by
          \DKLB ,\DFKZ
$$
\alpha_{i}^{-1}(M_{X}^{2}) =  k_{i} \alpha_{X} +{{b_i}\over {4\pi }}
\log {{M_{string}^2}\over {M_X^2}} - {1\over {4\pi }}
\sum _mb{'}_i^m ln(2ReT_m \mid \eta (T_m)\mid ^{4})
\eqn\stringthres
$$
 where $\eta (T)$ is the Dedekind function which admits a large $T$
expansion $\eta (T)\simeq e^{-\pi T/12}(1-e^{-2\pi T}+...)$.
The $T_m$ are the three complex scalars (untwisted moduli)
whose vevs give the mass scale of compactification.
                      Here $M_{string}$              is given by
\KAPLU
$$
M_{string}\ \simeq \ 0.7 \times g \times 10^{18} \ GeV
\eqn\mss
$$
and the constants $b{'}_i^m $      depend on the particular
orbifold model.
We see from this that string theories are $potentially$ more
predictive than Grand Unified theories for $M_{X}$ is not a free
parameter; we will discuss this point      shortly.

The minimal unification assumption has
                                    no significant threshold corrections
due to states at $M_{X}$. However there are threshold corrections at
the supersymmetry breaking scale that should be included in the
analysis for, as we discussed in section 3,         the
supersymmetric spectrum is not expected to be the degenerate one
assumed in obtaining Fig. 6.           By integrating the
renormalisation group equations              it is easy to
calculate the full spectrum as discussed above               and
to include it in the analysis of gauge coupling running.
\vskip 5.truein
   {\bf Fig.7}    The solid lines show the contours in the
$m_0,m_{1/2}$ plane corresponding to the value of $\alpha _s(M_Z)$
needed for unification of the gauge couplings. The dotted lines
show the $m_t$ contours needed to give the correct electroweak
breaking.
\endpage

 Once one has a mass spectrum the renormalisation group equations
for the gauge couplings may be integrated {\it up} in energy
using the experimentally determined values at $M_{Z}$ as boundary
values. The beta functions change at the appropriate mass scales
as the threshold for the supersymmetric states is passed and this
can be done in detail using the results of the renormalisation
group
      evaluation of the mass spectrum   by including a particular
supersymmetric contribution to the $\beta$ functions only at
scales above its mass. The result of this
analysis \RR\ is shown in Fig. 7 for $\mid \mu _0/m_0\mid =
0.2,0.4,1.0$ and $5.0$  (figs. 7-(a),(b),(c),(d) ).

           From the figure  it may be seen that the present
measurement of the strong coupling constant in the range 0.108-0.118
\MARTI\  , is consistent with
the unification prediction for a range of the supersymmetry
breaking masses $m_{1/2}        $ and $m_{o}$ between 2 and 90
Tev
          (the results shown correspond to the case
$A_0=B_0=0$).
           The sensitivity to the value taken for $\mu_{0}$ is
such that if it is reduced the mass scale for the remaining
supersymmetric states is increased. Broadly the result of
including the non-degenerate spectrum is to increase the
effective SUSY scale (the average scale of the SUSY breaking
masses) by a factor of 3-10 compared to the
analysis of Amaldi et al. \AMALDI\ the range corresponding to whether the
higgsino is light or not \RR\ .

It is perhaps appropriate to comment about the meaning of this
fit, for the three gauge couplings are described in terms of
three effective parameters $M_{X}$, $\alpha_{X}$ and the
effective supersymmetry mass scale, meaning there will always be
a fit and apparently no test of unification! However this is not
quite fair for the resulting values of the parameters must be
reasonable if the scheme is to make sense. Thus $M_{X}$ should
be less than the Planck scale but large enough to inhibit proton
decay in Grand Unified theories. Also $\alpha_{X}$ must be
positive and within the perturbative domain (although it may be
sensible to contemplate non-perturbative unification). Finally
the supersymmetry mass scale must be large enough to explain why
no supersymmetric states have been found and small enough to
avoid the hierarchy problem (as we will see the latter gives a
very strong constraint). In the analysis presented above the
values of the parameters satisfy these conditions consistent with
the hypothesis of minimal unification.

\endpage

\leftline{\bf 4.2\                The electroweak breaking scale}
\bigskip

We now turn to the calculation of spontaneous breaking of
electroweak symmetry by radiative effects. As discussed in
section 3         in the MSSM the Higgs masses are also
determined in terms of the supersymmetry breaking terms at the
unification scale . For a viable theory it is necessary that the
electroweak symmetry be spontaneously broken and the way that may
come about is for the radiative corrections to drive the Higgs
mass {\it squared} negative thus triggering spontaneous symmetry
breaking. As noted above the gauge interactions increase the
masses (squared) and only the (top) Yukawa interactions can drive
the mass squared negative. The effect of the radiative
corrections involving this coupling is to reduce the stop and the
Higgs masses squared but due to the large positive QCD radiative
corrections which affect only the stop it is the Higgs scalar
mass squared that is driven negative as desired. Since the
effective potential of the Higgs scalar is completely determined
in a supersymmetric theory by its gauge and Yukawa couplings its
resultant vacuum expectation value is fixed, corresponding to a
prediction for $M_W$ in terms of the parameters of
the model.              The most significant radiative correction
in this respect comes from the top Yukawa coupling so the value
of $M_{W}$ is largely determined by the  value of  $m_{t}$
\footnote*{
In determining the structure of electroweak breaking we choose
to start with the case $A_{0}=B_{0}=0$. We do this for simplicity
and also because it is also true in some string derived models.
                                                   Despite being
originally zero A and B are driven non-zero at scales below
$M_{X}$.}.

The Higgs potential which determines the scale of electroweak
breaking is given by eq.\pot  .  Since, cf eq. \semin\ , $\mu _2$ depends
on $h_t$, we see                 from eq.\const\
that tuning $h_0\equiv h_t(M_X) $ we may always obtain the correct value
of $M_Z$. Thus to each point of the solution plane of Fig.7  it is
possible to assign a definite $h_{0}$ (or equivalently $m_{0}$)
which gives the correct electroweak breaking scale. Using this
the results of the analysis of electroweak breaking may be
conveniently summarised by drawing contour plots of constant
$m_t$ (needed to give the correct $M_Z$)
        on the $m_{0},m_{1/2}$ plane.   These are shown \RR\ by the
dotted lines in Fig.7              for the case $m_{t}=100,160$ GeV
from which it may be seen that part of the previously allowed
region is excluded by the requirement of acceptable radiative
electroweak breaking coupled with the LEP bounds on the top mass.
(The solution has a potential bounded from below and ,     as
discussed in section 3,    colour and charge remain unbroken).
The                        origin of the variation with respect
to $m_0$ shown in fig. 7 may be easily understood because
                          larger $m_{0}$ requires larger $h_0$ to
drive $\mu _{2}^{2}$ negative at the correct scale. The same applies
to increasing the value of $\mu$.

 It may be seen from
Fig. 7             that a heavy top quark leads to a very
satisfactory explanation of both the existence of electroweak
breaking and its magnitude for a wide range of supersymmetry
breaking parameters. This provides a realisation of the structure
discussed in section 3, namely that in the MSSM electroweak breaking
is naturally triggered by radiative corrections.
\bigskip

\leftline{\bf 4.3 The fine tuning problem}

\bigskip

It is remarkable that the intricacies of this radiatively
generated mass spectrum are such that an acceptable scale of
electroweak breaking results for values for $m_{t}$ within the
range allowed by the precision LEP measurements. However there
is a concealed fine tuning problem associated with this solution.
The problem is that the value of the Higgs mass and hence of
$M_{W}$ is sensitively dependent on the top Yukawa coupling. For
example with the choice of parameters corresponding to
Fig. 7,             if $M_{W}$ is constrained within 0.4\%
(experimental uncertainty) then $h_{0}$ must be tuned to within
about two parts per million, or the top mass must be constrained
to about 0.3MeV! The reason for this extreme
sensitivity may be seen from the form of the term in the
renormalisation group responsible for the Higgs mass $\mu _2^2$,
(see eq.(43)       in the     Appendix).
 As we have discussed the last term proportional to the top
Yukawa coupling, $h$, drives the mass squared negative at some
point $Q_{0}$ far from $M_{X}$. Expanding around this point and,
for illustration, keeping  only the dominant term involving the
stop mass, below $Q_{0}$ the Higgs mass squared is given by (here
we also neglect the running of $\mu$)
$$
\mu _{2}^{2}(Q^{2})={3\over {16\pi^{2}}}h^{2}(m_{\tilde{t}}^{2}
+m_{\tilde{t}^{c}}^{2})log(Q^{2}/Q_{0}^{2})  \ .
\eqn\mastun
$$

It may be seen that the negative Higgs mass squared which sets
the scale for the vacuum expectation value of $H_{2}$ and hence
$M_{W}$ is proportional to the stop mass squared and to the top
Yukawa coupling squared. For larger squark masses $M_W$ will be
very sensitive to the value of the top Yukawa coupling, giving
rise to the fine-tuning problem noted above.

In ref. \BG\  ``reasonable''
scales for the
supersymmetry thresholds were estimated by demanding that the
sensitivity of the electroweak breaking scale to any of the
parameters of the standard model should be less than some value,
c. In the case we are interested in c is defined through

$$
{{    \delta M_{W}^{2}}\over {M_{W}^{2}}}=c{{\delta h^2}\over {h^2}}
\eqn\tuning
$$
where the scale is the electroweak breaking scale.

No fine tuning would correspond to $c\sim    1$ but, more
conservatively, a value of $c=10$ was chosen as a measure of a
reasonable theory {\it i.e.h} must be tuned to finer than
$1/10$         the W mass {\it uncertainty}.

Using this measure of the fine tuning problem we may determine
the implications for the supersymmetry breaking scale.
Ignoring the gauge couplings in the renormalization group equation for
$\mu _2^2$        and also the running of the squark and Yukawa
coupling one may solve eqs.    \mastun      and         \tuning
to find \RR
$$
c\sim
{    {ln(M_{X}^{2}/Q^{2})}\over {ln(Q_{0}^{2}/Q^{2})} } \ .
\eqn\ccc
$$

The enhancement of
$(      ln(M_{X}^{2}/Q^{2}) /ln(Q_{0}^{2}/Q^{2}) )$ in the
sensitivity of the W mass to the top Yukawa coupling comes from
the fact that $Q_{0}$ is driven by a large logarithm involving
the unification scale $M_{X}$. Although eq.\ccc             is
only a
       rough approximation it does serve to explain the origin
of the extreme sensitivity of the Higgs mass to the top Yukawa
coupling which may be found exactly
from the full renormalisation
group equations.  This is shown \RR\ in Fig.8
                                      using
instead of the approximate form of eq.\ccc   ,   the value of c
determined
           from the full renormalisation group equations.
\endpage
{}.

\vskip 6.truein
  {\bf     Fig. 8}  The solid lines give the contours of constant c
in the $m_0,m_{1/2}$ plane for $\mid \mu _0/m_0\mid =1, A_0=B_0=0$.
                Also shown are the $m_b$ contours (dashed lines)
following from the unification prediction $m_b(M_X)=m_{\tau }(M_X)$.
The dotted lines are as in fig. 7.

\endpage

It is found  that the condition $c\le 10$ restricts us to the region
$\mid      \mu_{0}/{m_{0}}\mid\sim     1$, $\alpha_{s}=0.118$ in
the $m_{0},m_{1/2}        $ plane. The full SUSY spectra for each
of the ``extremes '' labelled by Z and X in Fig.8             are
listed in Table 2 including the (small) contributions to the
masses coming from the spontaneous breaking of electroweak
symmetry eq.\masillas . The neutralino and chargino masses
 (labeled by their dominant content)
are obtained by diagonalising the full mass matrices. The spectra are
very strongly constrained and lie not far from the present
experimental limits with all SUSY masses less than 1TeV and
sleptons much less, of order
200GeV.  Note that the fine tuning constraint is largely
independent of the details of the
unification at $M_{X}$ and may be expected to give similar limits
on the SUSY spectrum in any unification scheme having a very
large $M_{X}$.

We have seen that within the minimal unification scheme the MSSM
with radiative corrections included is very predictive and that
the predictions are in excellent agreement with experiment
provided the supersymmetry breaking scale is low, with the new
supersymmetric states within reach of the LHC and SSC colliders.
As such the theory should be tested by these machines. In fact
one can make further predictions for one can also determine the
mass of the Higgs bosons from the results of this analysis. This
requires minimisation of the full Higgs potential and
determination of the masses at the minimum. For the parameter
choice of Fig.8             at the extremes Z and X the resulting
Higgs spectrum is given \RR\ in Table 2. However we should emphasise
that these masses are sensitive to the value of $tan\beta$, cf
eq.\higgs\   , and hence to the initial values $A_{0},B_{0}$.
As has recently been noted \HAHE\         there are also large
radiative corrections to the quartic terms in the effective
potential for large values of the top mass which increase the
light scalar mass, but these have been evaluated and are easily
included.

 The minimal unification of the gauge interactions also
determines the unification scale which we find to be
$M_{X}\approx 10^{16}Gev$, sufficient in Grand Unified
supersymmetric theories to inhibit nucleon decay (through dimension 6
operators) below
experimental limits. From the most optimistic point of view this
may be seen as evidence (albeit circumstantial) for unification.
Is there a realistic (string) model realising this simple scheme?
In Grand Unification the minimal

\begintable
 \   \|   {\bf Parameters}   |     \     \cr
$m_{1/2}$    \| 140    |   230    \cr
$m_0$   \|  190 |      120    \cr
$\mu _0$  \|   190     |     -120    \cr
$m_t$      \|     160     |    100   \cr
$tan\beta $      \|       21    |     5       \cr
  \    \|     {\bf Gauginos }     |  \             \cr
${\tilde {\gamma }}$   \|     57     |      83     \cr
${\tilde Z};{\tilde W}$    \|   99\ ,\ 99      |  120\ ,\ 112   \cr
${\tilde g} $    \|    354       |     559   \cr
 \      \|  {\bf Sleptons}   |    \    \cr
${\tilde L} $   \|    220     |     206       \cr
${\tilde E}_R$      \|    195  |  146     \cr
\   \|   {\bf Squarks}     |   \    \cr
${\tilde u}_L,{\tilde c}_L;{\tilde d}_L,{\tilde s}_L$\|365\ ;\ 373 |
511\ ;\  517   \cr
${\tilde u}_R,{\tilde c}_R$   \|   359    |   495    \cr
${\tilde d}_R,{\tilde s}_R,{\tilde b}_R$   \|   358        |  491   \cr
${\tilde t}_L;{\tilde b}_L$    \|   325 \ ,\ 335 | 491\ ,\ 497    \cr
${\tilde t}_R$    \|     273   |    452    \cr
    \   \|    {\bf Higgs, Higgsinos}      |    \     \cr
$h,H$      \|       91\ ,\ 264      |    84\ ,\ 221     \cr
$H^{\pm };P$    \|   276\ ,\ 264    | 232 \ ,\ 218    \cr
${\tilde H}^0$     \|   205\ ,\ 225   | 139\ ,\ 226    \cr
${\tilde H}^{\pm }$    \|   229    |     227
\endtable
\bigskip

\centerline{\bf  Table 2}
\endpage

\noindent
unification assumptions which
lead to the successful predictions discussed above can be
realised if the MSSM is embedded in a GUT such as SU(5) which
breaks directly to $SU(3)\otimes SU(2) \otimes U(1)$ at the
unification scale $M_{X}$ and all the resulting massive states
are nearly degenerate. There is no constraint in Grand
Unification which prohibits such a pattern of unification and it
provides a nice example of the minimal unification discussed
here.
In string theories, however, there are much stricter
constraints for a given four dimensional string theory has
definite multiplet structure. In principle one can construct a string
model with a unification group like $SU(5)$ or $SO(10)$ although
this requires the presence of higher Kac-Moody algebras \HIGH\
which is technically complicated \HIGHK \ . This type of scenario
would lead to results similar to any grand unification scheme.
On the other hand
                    the simplest realisation of minimal string
unification is for the gauge group after compactification to be
just $SU(3)\otimes SU(2)\otimes U(1)$ with the minimal particle
content {\it i.e.} there is no need for any additional heavy
states. While no example of such a string theory has yet been
constructed it seems likely that something quite close to it can
be found. However the prediction for $M_{X}$ is difficult to
satisfy for a recent survey \ILR ,\IL\
                                 of all abelian orbifold models
has revealed that large classes of models do not admit this
possibility ( although it is not excluded).
The alternative, \ANTON ,\IL\                  is to give up the
minimal unification assumption and to add states beyond the
minimal set. Although this is certainly a possibility one loses
the simplicity of the minimal model.

To summarize  the main conclusions of this section,
 we have examined above     the possibility of generating
electroweak breaking via radiative corrections.  The prediction
is sensitive to the top quark mass and it is in remarkable
agreement with experiment for a top quark mass in the range
needed for consistency with the precision LEP experiments.
However if the SUSY masses are near their upper value allowed by
the coupling unification the top quark mass must be fine tuned
to a very high degree {\it i.e.} the natural scale of W mass lies
close to the supersymmetric mass. This fine tuning may be reduced
only if the SUSY mass scale is low and asking that it be less
than one part in ten ($C<10$) forces
$m_{0},m_{1/2}        <\sim    10^{2}GeV$.
In this case consistency with the prediction for the strong
coupling constant is only possible for $\alpha_{s}$ at its
upper allowed value $\sim 0.118$ as determined by jet analyses and LEP
\MARTI\ .

\bigskip

\leftline {\bf 5\ Outlook}
\bigskip
We have discussed in the present  article the mechanism of
electroweak  symmetry breaking in the supersymmetric standard
model. It is quite remarkable   that the very structure of
the MSSM is such that quantum loop corrections induce in a
natural             way the $SU(2)_L\otimes U(1)$ symmetry
breaking. We think it is fair to say that the electroweak
symmetry breaking in the MSSM is much more satisfactory than
that in the standard model, not only because of the solution
that supersymmetry provides to the hierarchy problem but also
for the natural way in which this mechanism takes place.
Whereas in the minimal standard model one puts by hand a
         tachyonic mass for the Higgs field in order to generate
the desired minimum, in the supersymmetric model  this is
just a consequence of the particle content of the model and
the radiative corrections. The excellent agreement of the
renormalization group running of the coupling constants
with the measured values of $sin^2\theta _W$ is a further
support in favour of the presence of low energy supersymmetry.
Although some points remain unclear (e.g., the origin of
                                           supersymmetry breaking)
the above mentioned successes of the MSSM do not seem to
depend heavily on details but on the general structure of the
low energy model.

A very important property of low energy supersymmetry is that
it should be possible to test it in forthcoming accelerators.
We described above how indeed a relatively low mass spectrum
is expected if we want to avoid fine-tuning.
Thus the sparticles should be accesible to experimental production
at LHC and SSC. If this is the case   very exciting physics
is waiting to be discovered using     these machines.

\endpage

\centerline{\bf Appendix}

We collect here the renormalization group equations for the
couplings and parameters discussed in the main text.
We assume here the minimal particle content of the SSM
and neglect all Yukawa couplings except for that of the
top quark. The more general case with non-vanishing
tau     and bottom Yukawa couplings may be found e.g. in
ref. \IBLO ,\ILM . One has for the gauge couplings
$$
{{dg_i^2}\over dt}\ =\ -{{b_i}\over {(4\pi )^2}}g_i^4
\eqn \ggg
$$
and for the gaugino masses
$$
{{dM_i}\over {dt}}\ =\ -{{b_i}\over {(4\pi )^2}}g_i^2M_i
\eqn \mmm
$$
where $t\equiv log(M_X^2/Q^2)$  and  $b_3=-3$, $b_2=+1$
and $b_1=+11$ for the minimal particle content of the SSM.
The supersymmetric mass of the Higgsinos evolve according to the
equation
$$
{{d\mu ^2}\over {dt}}\ =\ \mu ^2\ (3{\tilde {\alpha }}_2
+{\tilde {\alpha }}_1-3Y_t)
\eqn\inos
$$
where one defines
$${\tilde {\alpha }}_i\ \equiv \ {{{\alpha }_i}\over {4\pi }}
\ ;\ Y_t\ \equiv \ {{h_t^2}\over {(4\pi )^2}}
\eqn \tild
$$
where $h_t$ is the Yukawa coupling of the top quark.
The Yukawa coupling of the third generation particles
have renormalization group equations
$$
\eqalign{
{{dY_t}\over {dt}}\ &=\ Y_t({{16}\over 3}{\tilde {\alpha }}_3+
3{\tilde {\alpha }}_2+{{13}\over 9}{\tilde {\alpha }}_1-6Y_t) \cr
{{dY_b}\over {dt}}\ &=\ Y_b({{16}\over 3}{\tilde {\alpha }}_3
+3{\tilde {\alpha }}_2+{7\over 9}{\tilde {\alpha }}_1- Y_t ) \cr
{{dY_{\tau }}\over {dt}}\ &=\ Y_{\tau }(3{\tilde {\alpha }}_2
+3{\tilde {\alpha }}_1)\ . \cr }
\eqn \yuk
$$
The equivalent equations for the Yukawas of the first two
generations
are obtained by deleting the $Y_t$ factor inside the brackets
except
for the Yukawas for the u and c quarks in which the factor six
is replaced by a three. The running of the masses of squarks and
sleptons are given by
$$
\eqalign{
{{dm_Q^2}\over {dt}}\ &=\ ({{16}\over 3}{\tilde {\alpha }}_3M_3^2
+3{\tilde {\alpha }}_2M_2^2+{1\over 9}{\tilde {\alpha }}_1M_1^2)
\ -\ Y_t(m_Q^2+m_U^2+\mu _2^2+A_U^2   -\mu ^2) \cr
{{dm_U^2}\over {dt}}\ &=\ ({{16}\over 3}{\tilde {\alpha
}}_3M_3^2+
{{16}\over 9}{\tilde {\alpha }}_1M_1^2)\ -\ 2Y_t(m_Q^2+m_U^2+
\mu _2^2+A_t^2   -\mu ^2)    \cr
{{dm_D^2}\over {dt}}\ &=\ ({{16}\over 3}{\tilde {\alpha
}}_3M_3^2+
{4\over 9}{\tilde {\alpha }}_1M_1^2)  \cr
{{dm_L^2}\over {dt}}\ &=\ (3{\tilde {\alpha }}_2M_2^2
{\tilde {\alpha }}_1M_1^2)   \cr
{{dm_E^2}\over {dt}}\ &=\ (4{\tilde {\alpha }}_1)\ .   \cr }
\eqn \mass
$$
For the scalars of the first two generations the same equations
apply with $Y_t=0$. Concerning the Higgs doublets mass parameters
one finds
$$
\eqalign{
{{d\mu _1^2}\over {dt}}\ &=\ (3{\tilde {\alpha }}_2M_2^2+
{ \tilde {\alpha }}_1M_1^2)\ +\ (3{ \tilde {\alpha }}_2+
{ \tilde {\alpha }}_1)\mu ^2\ -\ 3Y_t\mu ^2  \cr
{{d\mu _2^2}\over {dt}}\ &=\ (3{ \tilde {\alpha }}_2M_2^2+
{ \tilde {\alpha }}_1M_1^2)\ +\ (3{ \tilde {\alpha }}_2+{ \tilde
{\alpha
}}_1)\mu ^2\ -\ 3Y_t(m_Q^2+m_U^2+\mu _2^2+A_t^2   )  \cr
{{d\mu _3^2}\over {dt}}\ &=\ ({3\over 2}{ \tilde {\alpha }}_2
+{1\over 2}{\tilde {\alpha }}_1-{3\over 2}Y_t)\mu _3^2\ +\
3\mu  A_tY_t\ -\  \mu (3{\tilde {\alpha }}_2M_2+{\tilde {\alpha
}}_1
M_1)\ . \cr }
\eqn \mahiggs
$$
Finally, the trilinear soft terms associated to the third
generation Yukawa couplings are
$$
\eqalign{
{{dA_t}\over {dt}}\ &=\ ({{16}\over 3}{\tilde {\alpha }}_3
  M_3         +3{\tilde {\alpha }}_2  M_2         +
{{13}\over 9}  M_1         )\ -\  6Y_tA_t  \cr
{{dA_b}\over {dt}}\ &=\ ({{16}\over 3}{\tilde {\alpha }}_3
{{M_3}       }+3{\tilde {\alpha }}_2{{M_2}       }+
{7\over 9}{\tilde {\alpha }}_1  M_1         )\ -\ Y_tA_t  \cr
{{dA_{\tau }}\over {dt}}\ &=\ (3{\tilde {\alpha }}_2
{{M_2}       }+3{\tilde {\alpha }}_1{{M_1}       })\ .  \cr }
\eqn \aaa
$$
For the soft terms associated to the first two generations
the same equations apply setting $Y_t=0$ except for the
soft terms $A_u$ and $A_c$ in which one has to replace
the factor six by one in the first of eq. \aaa  .

The functions  $E_i$ and $F_1$ appearing in eq.\renyuk
are given by
$$
\eqalign{
E_1(t)\ &=\ (1+b_3{{\alpha _3(0)}\over {4\pi }}t)^{16/3b_3}
(1+b_2{{\alpha _2(0)}\over {4\pi }}t)^{3/b_2}
(1+b_1{{\alpha _1(0)}\over {4\pi }}t)^{13/9b_1} \cr
E_2(t)\ &=\ (1+b_1{{\alpha _1(0)}\over {4\pi }}t)^{-2/3b_1}
E_1(t)  \cr
E_3(t)\ &=\ (1+b_2{{\alpha _2(0)}\over {4\pi }}t)^{3/b_2}
(1+b_1{{\alpha _1(0)}\over {4\pi }}t)^{3/b1}  \cr
F_1(t)\ &=\ \int _0^t E_1(t')dt'   \ , \cr
}
\eqn \eee
$$
where $b_i$ are the coefficients of the three $\beta $-functions.

\bigskip
\bigskip
\bigskip
\bigskip

 \par \penalty-400 \vskip\chapterskip
   \spacecheck\referenceminspace \immediate\closeout\referencewrite
   \referenceopenfalse
   \line{\fourteenrm\hfil REFERENCES\hfil}\vskip\headskip
   \input referenc.texauxil
   

\vfill\eject\bye